\documentclass[11pt]{article} 
\usepackage[utf8]{inputenc}
\usepackage[left=1in,right=1in,top=1in,bottom=1in]{geometry}
\usepackage{xcolor}
\usepackage{url}
\usepackage{cite}
\usepackage{graphicx}
\usepackage{wrapfig}
\usepackage{authblk}
\usepackage{hyperref}
\usepackage{algorithm}
\usepackage{algpseudocode}
\usepackage{longtable}
\hypersetup{
    colorlinks=true,
    linkcolor=blue,
    filecolor=magenta,
    urlcolor=cyan,
    citecolor=black,
    pdftitle={Overleaf Example},
    pdfpagemode=FullScreen,
    }
\usepackage{amsmath,amsfonts,amsthm,bm}
\usepackage{subcaption}

\usepackage{graphicx} 

\usepackage{siunitx}
\DeclareSIUnit\angstrom{\textup{\AA}}

\usepackage{graphicx}
\usepackage{tikz}
\usetikzlibrary{arrows.meta}
\usetikzlibrary{datavisualization}
\usetikzlibrary{datavisualization.formats.functions}

\usepackage{comment}

\newcommand{\eqn}[1]{Eq.~(\ref{#1})}

\newcommand{\sect}[1]{Section~\ref{#1}}
\newcommand{\sects}[1]{Sections~\ref{#1}}
\newcommand{\sectx}[1]{\ref{#1}}
\newcommand{\app}[1]{Appendix~\ref{#1}}

\newcommand{\fig}[1]{Fig.~\ref{#1}}

\newcommand{\tab}[1]{Table~\ref{#1}}
\newcommand{\alg}[1]{Algorithm~\ref{#1}}

\title{Uncertainty-quantified $J$-integral computation for quasicontinuum and finite element methods}

\author[1]{Sai Harshith Badi}
\affil[1]{
Department of Aerospace Engineering and Mechanics,
University of Minnesota, Minneapolis, MN 55455, USA}
\author[1]{Stephen M. Whalen}
\author[2]{Ronald E. Miller}
\affil[2]{
Department of Mechanical and Aerospace Engineering,
Carleton University, Ottawa, Canada}
\author[1]{Ellad B. Tadmor\thanks{Contact author: tadmor@umn.edu}}

\newcommand{\cC} { \ensuremath{\mathcal C}}
\newcommand{\cN} { \ensuremath{\mathcal N}}

\newcommand{\bff} { \ensuremath{\boldsymbol {f}}}
\newcommand{\bn} { \ensuremath{\boldsymbol {n}}}
\newcommand{\bq} { \ensuremath{\boldsymbol {q}}}
\newcommand{\br} { \ensuremath{\boldsymbol {r}}}
\newcommand{\bu} { \ensuremath{\boldsymbol {u}}}
\newcommand{\bx} { \ensuremath{\boldsymbol {x}}}

\newcommand{\bF} { \ensuremath{\boldsymbol {F}}}
\newcommand{\bH} { \ensuremath{\boldsymbol {H}}}
\newcommand{\bI} { \ensuremath{\boldsymbol {I}}}
\newcommand{\bN} { \ensuremath{\boldsymbol {N}}}
\newcommand{\bP} { \ensuremath{\boldsymbol {P}}}
\newcommand{\bQ} { \ensuremath{\boldsymbol {Q}}}
\newcommand{\bT} { \ensuremath{\boldsymbol {T}}}
\newcommand{\bX} { \ensuremath{\boldsymbol {X}}}

\newcommand{\bpsi} {\ensuremath{\boldsymbol { \psi }}}

\newcommand{\rI} { \ensuremath{\text {I}}}

\date{}

\graphicspath{{figures/}{./}}

\begin{document}

\maketitle

\begin{abstract}

The $J$-integral is a fundamental concept in fracture mechanics, quantifying the energy release rate that drives crack propagation. 
This work presents a rigorous implementation and validation of the $J$-integral within the multiscale three-dimensional (3D) quasicontinuum (QC3D) method along with an uncertainty quantification analysis applicable to both QC3D and conventional FE methods. We compute the 3D $J$-integral under plane-strain assumptions, utilizing continuum mechanics fields (stress and strain energy density) derived from the underlying interatomic potential through the Cauchy--Born rule. Our implementation is comprehensively validated against both linear elastic fracture mechanics (LEFM) theory and the virtual crack extension (VCE) method for three regimes of increasing complexity: 
small strain elasticity, and nonlinear deformation without and with atomic relaxation in the region near the crack tip.
Excellent agreement is demonstrated in all cases. Further, we introduce a novel Markov chain Monte Carlo (MCMC) framework to statistically assess the uncertainty in $J$-integral results for a given mesh and integration domain in QC3D and conventional FE methods. Our implementation's predictive capability is demonstrated through a QC3D simulation of a three-point bending test on silicon, where the computed critical energy release rate is shown to be in excellent agreement with the Griffith criterion. This work establishes a reliable and efficient framework for evaluating crack driving forces in multiscale fracture simulations with quantified uncertainty, enabling large-scale fracture simulations while resolving atomistic mechanisms at the crack tip.
\end{abstract}

\section*{Nomenclature}
\begingroup
\small
\begin{longtable}{@{}p{0.22\textwidth}p{0.74\textwidth}@{}}
\multicolumn{2}{@{}l}{\textit{Latin symbols}}\\[2pt]
$a$ & Crack length \\
$a(\mathbf{q}',\mathbf{q})$ & Metropolis--Hastings acceptance probability \\
$\mathbf{a}_i$ & Deformed lattice vectors \\
$A$ & Atomistic (nonlocal) region; crack area; surface area in the surface-energy calculation \\
$A_0$ & Integration domain (area) in the reference configuration \\
$\mathbf{A}_i$ & Reference lattice vectors \\
$B$ & Thickness in the out-of-plane (crack-front) direction \\
$C$ & Continuum (local) region; closed contour enclosing $A_0$ \\
$C_0$, $C_\pm$ & Outer contour; upper and lower crack-face segments \\
$\cC$ & Space of admissible configurations \\
$d$ & Initial crack opening \\
$E$, $E'$ & Young's modulus; effective modulus \\
$E^\alpha$ & Energy of atom $\alpha$ \\
$E_{\rm bulk}$, $E_{\rm slab}$ & Relaxed energies of the periodic crystal and of the slab with free surfaces \\
$E_{\rm cell}$ & Energy of a unit cell of the deformed perfect crystal \\
$f$ & Tolerance factor for cleavage detection \\
$\bff^{\rm ext,\beta}$ & External force on repatom $\beta$ \\
$\bff^{\rm ghost,\beta}$ & Ghost-force correction on repatom $\beta$ \\
$\bF$, $\bF^e$ & Deformation gradient; deformation gradient of element $e$ \\
$G$, $G_{\rm c}$ & Energy release rate; critical energy release rate \\
$h$ & Characteristic element size \\
$\bH$, $\bH^e$ & Eshelby (energy--momentum) tensor; Eshelby tensor of element $e$ \\
$\bI$ & Second-order identity tensor \\
$J$ & $J$-integral \\
$J^e$ & Contribution of element $e$ to the $J$-integral \\
$J^\Gamma$, $J^{A_0}$ & Contour and domain forms of the $J$-integral \\
$J_h^\Gamma$, $J_h^{A_0}$ & Contour and domain forms evaluated with the discrete fields \\
$J_{\rm c}$ & Critical value of $J$ at cleavage \\
$J^{(k)}$, $\bar J$ & Sampled $J$ at step $k$; mean of the sampled $J$ \\
$k$ & MCMC step index; lag \\
$K$, $K_\rI$ & Stress intensity factor; mode~I stress intensity factor \\
$K_{\rm c}$ & Fracture toughness (critical stress intensity factor) \\
$K_{\rI,0}$, $\Delta K_\rI$ & Initial value and increment of $K_\rI$ in the loading scheme \\
$M$ & Truncation lag of the autocorrelation sum \\
$\boldsymbol{M}$ & Unit outward normal to $A_0$ on $C$ \\
$\hat\bn$ & Unit vector in the crack-growth direction \\
$n_{\rm rep}$, $n_{\rm elements}$ & Number of repatoms; number of elements \\
$\bN$ & Unit outward normal to a contour or surface \\
$N^\alpha$, $N^\alpha_{,K}$ & Shape function of node $\alpha$; its derivative with respect to $X_K$ \\
$\cN^e$ & Set of nodes of element $e$ \\
$N_{\rm burn}$, $N_{\rm sample}$ & Number of burn-in steps; number of retained samples per chain \\
$N_{\rm c}$ & Number of MCMC chains \\
$N_{\rm e}$ & Number of elements in the integration domain \\
$N_{\rm eff}$ & Effective sample size \\
$N_{\rm int}$ & Number of interior nodes of the integration domain \\
$p(\mathbf{q})$ & Target distribution of the interior nodal $q$-values \\
$\bP$, $\bP^e$ & First Piola--Kirchhoff stress tensor; its value in element $e$ \\
$q$, $\bq$ & Scalar weighting ($q$-) field; vector weighting field $\bq = q\hat\bn$ \\
$q^\alpha$, $\mathbf{q}$ & Nodal value of $q$ at node $\alpha$; vector of interior nodal $q$-values \\
$q_{\rm lin}$ & Linearly interpolated $q$-field \\
$\bQ$ & Configurational force \\
$r$ & Distance from the crack tip; Metropolis--Hastings ratio \\
$\br$, $\br_A$, $\br_P$ & Positions of repatoms; of atoms in the atomistic region; of padding atoms \\
$R(q)$, $R_h(q)$ & Residual of the domain integral for exact and discrete fields \\
$\hat R$ & Gelman--Rubin statistic \\
$s(\mathbf{q}'\mid\mathbf{q})$ & Proposal distribution \\
$s_j^2$ & Sample variance of chain $j$ \\
$S$ & Surface enclosing a defect \\
$\bT$ & Referential traction vector \\
$\bu$, $\bu^\alpha$ & Displacement vector; displacement of node $\alpha$ \\
$v$ & Variance of an interior nodal $q$-value \\
$V_0$, $V^e$ & Integration volume in the reference configuration; volume of element $e$ \\
$V_{\rm within}$, $V_{\rm between}$, $\hat V$ & Within-chain, between-chain, and pooled variances \\
$w$ & Specimen width \\
$w^\alpha$ & Weight of node $\alpha$ in the linear form of $J$ \\
$W$, $W^e$ & Strain energy density; its value in element $e$ \\
$W_{\rm CB}$ & Cauchy--Born strain energy density \\
$\bX$ & Position vector in the reference configuration \\
$\bx_{\rm tip}$ & Crack-tip position \\
$x$, $y$, $z$ & Cartesian coordinates (crack-growth, crack-plane normal, and crack-front directions) \\[6pt]
\multicolumn{2}{@{}l}{\textit{Greek symbols}}\\[2pt]
$\alpha$, $\beta$ & Atom or node indices \\
$\gamma_{\rm s}$ & Surface energy \\
$\hat\gamma(k)$ & Estimated autocovariance at lag $k$ \\
$\Gamma$ & Contour surrounding the crack tip (inner contour) \\
$\delta$ & Deviation from the extrapolated energy (cleavage detection) \\
$\delta_{IJ}$ & Kronecker delta \\
$\Delta a$ & Virtual crack extension \\
$\kappa(\mathbf{q}'\mid\mathbf{q})$ & MCMC transition kernel \\
$\lambda$ & Probability \\
$\nu$ & Poisson's ratio \\
$\Pi$ & Total potential energy \\
$\Pi_1$, $\Pi_2$ & Energies of the reference and virtually extended configurations \\
$\rho(k)$, $\hat\rho(k)$ & Autocorrelation function; its estimate \\
$\boldsymbol{\sigma}$ & Cauchy stress tensor \\
$\sigma_J$ & $q$-field uncertainty (standard deviation of the sampled $J$) \\
$\tau$, $\hat\tau$ & Integrated autocorrelation time; its estimate \\
$\bpsi$ & Displacement field in the incremental loading scheme \\
$\Omega_0$ & Reference unit-cell volume \\[6pt] \multicolumn{2}{@{}l}{\textit{Subscripts and superscripts}}\\[2pt]
$(\cdot)_{\rm c}$ & Critical value \\
$(\cdot)^e$ & Element quantity \\
$(\cdot)_h$, $(\cdot)^h$ & Quantity computed from the discrete fields \\
$(\cdot)^{\rm init}$ & Initial guess \\
$(\cdot)_I$ & Mode~I \\
$(\cdot)_0$ & Reference configuration; initial value \\
$I, J, K$ & Indices referring to the reference configuration \\
$i, j, k$ & Indices referring to the deformed configuration \\
$(\cdot)_{,K}$ & Partial derivative with respect to $X_K$ \\
$(\cdot)^T$ & Transpose \\[6pt]
\multicolumn{2}{@{}l}{\textit{Abbreviations}}\\[2pt]
2D, 3D & Two-, three-dimensional \\
FE & Finite element \\
LEFM & Linear elastic fracture mechanics \\
MCMC & Markov chain Monte Carlo \\
QC, QC3D & Quasicontinuum; three-dimensional quasicontinuum \\
QOI & Quantity of interest \\
SIF & Stress intensity factor \\
UQ & Uncertainty quantification \\
VCE & Virtual crack extension \\\\
\end{longtable}
\endgroup

\section{Introduction}
\label{sec:intro}

The prediction of fracture associated with the initiation and propagation of cracks remains a central concern for the mechanics of solids, with critical implications for the safety and reliability of engineering structures across aerospace, civil, and mechanical engineering. Within the theory of fracture mechanics, an existing crack will propagate when the configurational force,\footnote{A generalized force that is conjugate to the crack tip position.} referred to as the \textit{energy release rate} $G$, reaches a critical value, $G_{\rm c}$, corresponding to the material's fracture toughness. Under linear elastic conditions, in which linear elastic fracture mechanics (LEFM) holds, $G$ is uniquely related to the stress intensity factor (SIF), $K$, which characterizes the stress state at the crack tip. In this case, fracture toughness is identified with the critical SIF, $K_{\rm c}$.  

A variety of techniques have been proposed to compute the energy release rate and predict the onset of fracture in numerical simulations, in particular for the finite element (FE) method. Banks-Sills \cite{banks-sills:1991} classified these techniques into \textit{direct} and \textit{indirect} methods, based on how they process the results. Direct methods extract $K$ by fitting the computed nodal displacements or stresses near the crack tip to the known singular asymptotic (Williams) crack-tip solutions, in which $K$ appears as the scaling coefficient of the leading $r^{-1/2}$ (stress) or $r^{1/2}$ (displacement) term. For example, the displacement extrapolation method \cite{chan:tuba:wilson:1970} uses the computed nodal displacements along the crack face at a sequence of increasing distances $r$ from the tip, fits these to the leading-order term of the Williams series solution, and extrapolates the fit to $r\to0$ to obtain $K$. The analogous stress extrapolation method \cite{barsoum:1976} performs the same procedure using computed stresses ahead of the crack tip instead of displacements. Because both approaches depend on how accurately the FE mesh reproduces this specific singular field shape in the immediate vicinity of the crack tip---a region that is difficult to resolve without specialized elements (e.g., quarter-point or singular elements)---their accuracy is sensitive to mesh refinement near the tip.

Alternatively, indirect methods are based on the fundamental definition of the energy release rate as the change in potential energy with increasing fracture surface area; examples include the stiffness derivative method \cite{parks:1974}, the $J$-integral method \cite{rice:1968}, the virtual crack extension (VCE) method \cite{hellen:1975}, the virtual crack-closure technique \cite{rybicki:kanninen:1977}, the global energy change method \cite{lindley:1992}, and the energy-derivative technique \cite{fan:benjar:2007}.\footnote{It is important to note that these methods are not entirely distinct; in many cases, they are fundamentally similar, merely offering different procedural avenues to achieve the same calculation.} Banks-Sills \cite{banks-sills:2010} reviewed these different FE methods for computing the energy release rate and demonstrated that energy-based methods tend to be the most accurate. Of these, the dominant approach implemented in FE packages is based on the $J$-integral method due to Rice \cite{rice:1968}, who showed that $G$ can be computed for both linear and nonlinear systems via a path-independent contour integral quantifying the energy flux to the crack tip. However, the direct implementation of the $J$-integral in FE as a path integral leads to poor accuracy; the calculated results exhibit path dependence due to numerical discretization errors, approximate integration, and modeling inaccuracies, violating the theoretical principle that the $J$-integral is based on. Instead, better results are obtained by recasting the $J$-integral as a domain (surface or volume) integral \cite{li:shih:1985,shih:moran:1986}. Several studies have compared the contour integral and domain integral approaches in FE and demonstrated the superiority of the latter \cite{banks-sills:sherman:1992}. For this reason, commercial FE packages such as ABAQUS and ANSYS internally implement the domain integral approach following the method of Shih et al.~\cite{shih:moran:1986}. Related work extended the domain integral in several directions. For mixed-mode and three-dimensional (3D) problems, Shivakumar et al.~\cite{shiva:raju:1992} generalized the equivalent domain integral formulation, while Yau et al.~\cite{yau:wang:corten:1980} and Wang et al.~\cite{wang:corten:yau:1980} introduced the interaction integral ($M$-integral) to extract individual mode-I, -II, and -III SIFs for isotropic and anisotropic materials, respectively---a technique now standard in commercial FE codes. For nonlinear material response, Simha et al.~\cite{simha:fischer:2008} developed $J^{\rm ep}$ to extend the domain integral beyond the proportional-loading restriction of conventional $J$ using incremental plasticity theory. For complex crack geometries, Okada and Ohata~\cite{okada:ohata:2013} extended the domain integral to non-planar, curved, and kinked 3D crack fronts.

While fracture mechanics is a continuum theory, crack propagation is fundamentally an atomistic process governed by the complex interplay of nanoscale mechanisms, including the breaking of atomic bonds, the nucleation and motion of dislocations, the evolution of nanoscale voids, among others. For this reason, atomistic simulations of fracture are of great interest for predicting and elucidating fracture behavior \cite{bitzek:kermode:2015}. In such simulations, material behavior is governed by an interatomic potential describing the bonding between atoms without the need for the phenomenological constitutive laws and failure criteria required by continuum-based FE methods.

To bridge the atomistic and continuum views of fracture, methods for computing the energy release rate at the atomic scale have been proposed. One of the earliest attempts was by Inoue et al.~\cite{inoue}, who computed an atomistic $J$-integral through direct summation over atoms on a closed contour surrounding a crack tip. Their method aggregated per-atom contributions of the potential energy, virial stress, and a locally estimated displacement gradient for each atom on the contour. Nakatani et al.~\cite{nakatani} proposed a domain integral approach to calculate the atomistic $J$-integral by weighting the contribution of atoms based on their position within the integration domain. The results agreed with LEFM at low stress intensity factors but deviated at high values due to geometric nonlinearities. Xu et al.~\cite{xu} computed $J$ using the fundamental definition of the energy release rate as a configurational force rather than contour or domain integration. They used the potential energy difference between two identically loaded atomistic models with cracks of length $a$ and $a + \Delta a$. However, no comparison with theory was made to validate the results. Jones et al.~\cite{jones} estimated the atomistic $J$-integral by applying Hardy's method \cite{hardy} to construct stress, displacement, and strain energy density fields from atomistic data in a manner consistent with the continuum balance laws of mass, momentum, and energy. They validated their results in the small-strain regime.

While atomistic simulations provide a detailed view of nanoscale crack-tip processes, fracture is fundamentally a multiscale phenomenon coupling far-field boundary conditions to crack propagation. Direct simulation of all atoms in a macroscopic system containing a crack is infeasible and also unnecessary, given that only a small fraction of atoms near a crack tip undergo displacements that vary significantly on atomic length scales, while the vast majority deform in a locally uniform fashion that is well described by continuum theory. This observation motivated the development of concurrent multiscale methods that seek to reduce the number of degrees of freedom by retaining full atomistic resolution only where necessary, such as at a crack tip, while treating the surrounding material with a computationally efficient continuum model \cite{tadmor:miller:2011}. An early example of this approach is the quasicontinuum (QC) method of Tadmor et al.~\cite{tadmor:ortiz:1996}. In QC, the domain is partitioned into an atomistic region (referred to as the ``nonlocal region''), where the energy is computed as a sum over the energies of atoms using a suitable interatomic potential, and a coarse-grained continuum region (``local region'') modeled using nonlinear FE with an atomistic constitutive model based on the Cauchy--Born rule. QC has been widely applied to study atomistic fracture processes in a variety of systems and loading conditions \cite{miller:tadmor:1998, miller:ortiz:1998, gao:klein:1998, gao:ji:1998, pillai:miller:2001, miller:tadmor:2002, zhang:klein:2002, hai:tadmor:2003, tadmor:hai:2003, gao:ji:2003, thiagarajan:hsia:2004, zhang:ge:2005, zhou:yang:2009, das:tirtom:2010, mei:li:2010, yu:shen:2010, shao:wang:2010, peng:lu:2011, vatne:ostby:2011a, vatne:ostby:2011b, mei:li:2011, lu:li:2011, shao:yang:2012, iacobellis:behdinan:2013, rokos:peerlings:2017, qiu:lin:2017, xu:fan:2017, qiu:lin:2018, ghareeb:elbanna:2020, wu:fang:2020a, wu:fang:2020b, ghareeb:elbanna:2021}.

The inherent suitability of multiscale methods, such as QC, for fracture simulations has generated significant interest, as they efficiently bridge the critical atomistic processes at a crack tip with the long-range constitutive response of the material. To fully leverage the predictive power of these frameworks for fracture simulations, a robust method for computing a fracture parameter is essential. This work addresses this need by implementing $J$-integral calculations within the QC framework. Specifically, the $J$-integral is computed by domain integration in the continuum region utilizing stress and strain energy density fields derived from the underlying atomistic description via the Cauchy--Born rule. The results are validated under controlled loading conditions against LEFM in the small-strain limit and, in the finite-strain limit, against the alternative VCE approach \cite{hellen:1975} for evaluating energy release rates. To quantify the uncertainty in the $J$-integral computations, we introduce a novel Markov chain Monte Carlo (MCMC) sampling technique over the weighting field $q$ used in the domain-integral form of $J$-integral. This approach is also demonstrated for conventional FE methods. The resulting formulation provides a reliable measure of the crack driving force in a multiscale setting, directly bridging atomistic fracture processes and continuum fracture mechanics.

The paper is organized as follows. \sect{sec:theory} provides a brief introduction to the $J$-integral and QC method. \sect{sec:jinqc} describes the implementation of the $J$-integral within the QC framework. In \sect{sec:validate}, the method is validated against LEFM and VCE. \sect{sec:sampling} introduces MCMC sampling for uncertainty quantification. The method is then applied to a three-point bending problem in \sect{sec:application}, and the paper concludes in \sect{sec:conclusions} with a summary and future research directions.

\section{Theoretical background}
\label{sec:theory}

\subsection{$J$-integral}
\label{sec:jint}

The mathematical description of defects in a continuum---be it an inclusion, a dislocation, or a crack---is a fundamental pursuit in solid mechanics. A pivotal advancement in this area was made by J.~D.~Eshelby in the 1950s through his work on the mechanics of defects \cite{eshelbyone, eshelbytensor}, which introduced the concept of the energy--momentum tensor, now known as the Eshelby stress tensor, $\bH$, given by
\begin{equation}
H_{IJ} = W \delta_{IJ} - P_{kJ} \frac{\partial u_k}{\partial X_I},
\label{eq:eshelby}
\end{equation}
where $W$ is the strain energy density, $\delta_{IJ}$ is the Kronecker delta, $\bP=\partial W/\partial \bF$ is the first Piola--Kirchhoff stress tensor, $\bF$ is the deformation gradient, $\bu$ is the displacement vector, and $\bX$ is the position vector in the reference configuration. We adopt the conventional continuum mechanics notation in which uppercase and lowercase indices refer to the reference and deformed configurations, respectively, and Einstein's summation convention is applied over repeated indices of the same case \cite{tadmor:miller:2012}.

The configurational force $\bQ$ on a defect follows as the integral of the divergence of the Eshelby tensor over a volume enclosing it, which by the divergence theorem can be expressed as a surface integral:
\begin{equation}
Q_I = \int_V \frac{\partial H_{IJ}}{\partial X_J}  dV = \int_S H_{IJ} N_J  dS,
\label{eq:eshelbysurf}
\end{equation}
where $S$ is any surface enclosing the defect and $\bN$ is the outward normal to that surface. This derivation provides a profound insight, namely that the force driving defect motion can be computed from a field evaluated on a remote surface enclosing it, independent of the complex details of the defect core itself.

\begin{figure}[t]\centering
     \begin{subfigure}[b]{0.3\textwidth}
         \centering
         \includegraphics[trim = 0mm 0mm 0mm 0mm, clip=true,width=\textwidth]{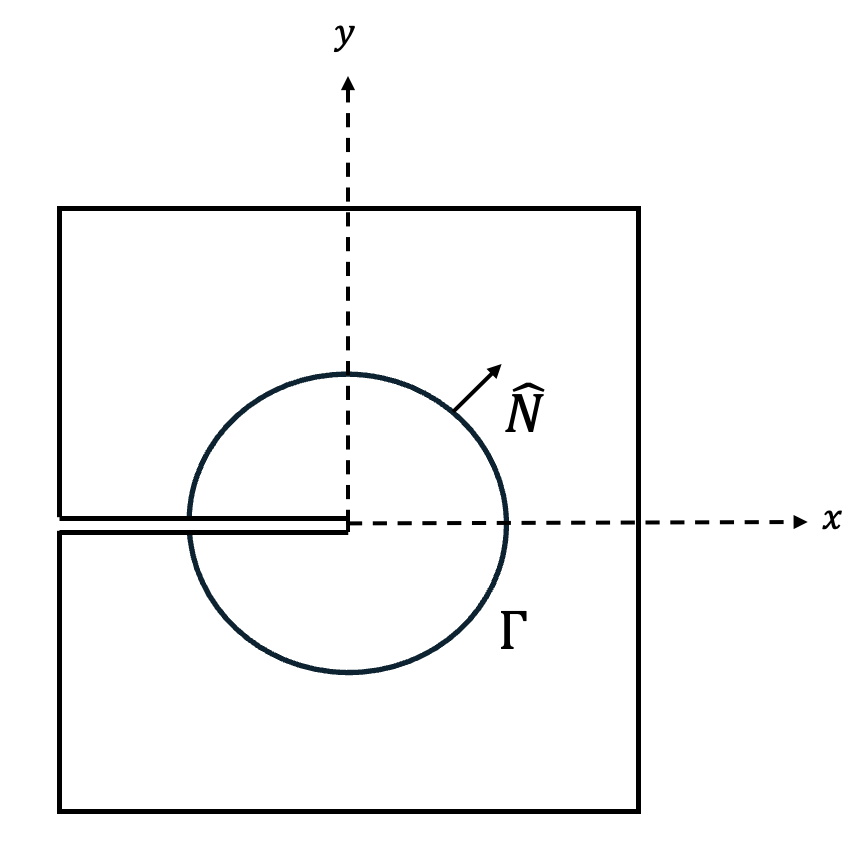}
         \caption{}
         \label{contourint}
     \end{subfigure}
     \begin{subfigure}[b]{0.3\textwidth}
         \centering
         \includegraphics[trim = 0mm 0mm 0mm 0mm, clip=true,width=\textwidth]{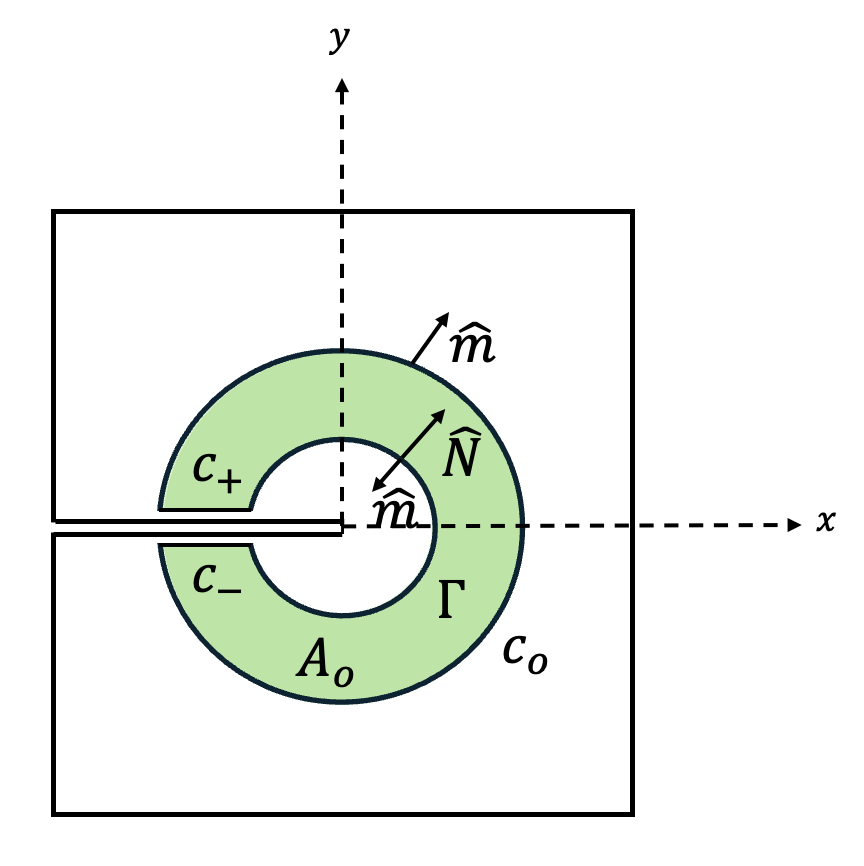}
         \caption{}
         \label{areaint}
     \end{subfigure}
     \begin{subfigure}[b]
     {0.3\textwidth}
         \centering
         \includegraphics[trim = 0mm 0mm 0mm 0mm, clip=true,width=\textwidth]{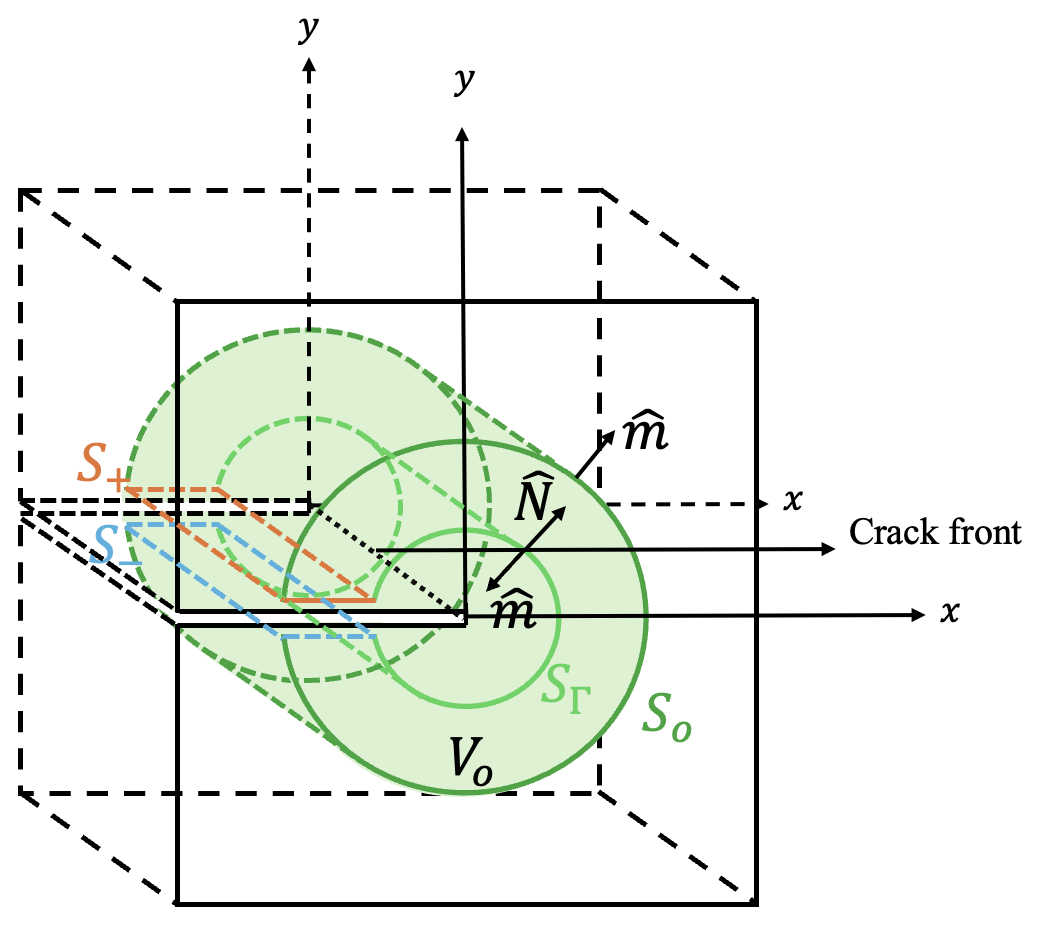}
         \caption{}
         \label{volumeint}
     \end{subfigure}
     \caption{(a) Schematic of the contour $\Gamma$ over which $J$ is evaluated in the 2D contour-integral form. (b) Closed contour $C = C_0 + C_+ - \Gamma+C_-$ enclosing the domain $A_0$ over which $J$ in the 2D domain (area) integral form is evaluated. (c) Closed surface $S=S_0+S_+-S_\Gamma+S_-$ enclosing volume $V_0$ over which $J$ in the 3D volume integral form is evaluated.}
     \label{}
\end{figure}

In a seminal paper \cite{rice:1968}, J.~R.~Rice applied Eshelby's framework to the analysis of cracks by introducing the two-dimensional (2D) $J$-integral formulation. Rice showed that for a straight crack lying along the $X_1$ direction inside a homogeneous, nonlinear elastic body under quasistatic conditions, the $X_1$ component of the configurational force on the crack tip is given by a contour integral $J$ defined as
\begin{equation}
J = \int_{\Gamma} \left( W N_1 - {T_i}  \frac{\partial {u_i}}{\partial X_1} \right) d\Gamma,
\label{eq:jcontour}
\end{equation}
where $\Gamma$ is any counterclockwise contour surrounding the crack tip as shown in \fig{contourint}, $\bN$ is the outward normal to the contour, and $\bT = \bP \cdot \bN$ is the referential traction vector. Rice proved that this contour integral is path-independent and equal to the energy release rate $G$ for crack advance,
\begin{equation}
J = G = -\frac{d\Pi}{dA},
\label{eq:err}
\end{equation}
where $\Pi$ is the total potential energy and $-d\Pi$ is its decrease due to an infinitesimal increase $dA$ in crack area.

For numerical implementation in an FE code, the 2D contour integral is typically converted into a 2D domain integral form (\fig{areaint}) through the application of the divergence theorem \cite{li:shih:1985}. This conversion introduces an arbitrary, continuous, piecewise-differentiable scalar weighting function $q(\bX)$ that equals unity on the inner contour and zero on the outer contour of the integration domain (see \app{app:contour-to-domain}), effectively spreading the contour evaluation over a finite domain,
\begin{equation}
J = \int_{A_0} \left[ P_{jK} \frac{\partial u_j}{\partial X_1} - W \delta_{1K} \right] \frac{\partial q}{\partial X_K}  dA_0,
\label{eq:jdomain2D}
\end{equation}
where $A_0$ is the integration domain in the reference configuration.

The standard 2D domain integral formulation can be generalized to a volumetric representation \cite{shih:moran:1986}, in which the integration is performed over a 3D domain enclosing a segment of the crack front. This formulation yields the total energy release rate for the enclosed segment. Here, we consider a straight crack in a 3D domain under plane-strain conditions. This enforces a state of deformation in which all cross-sections normal to the crack front are identical, as shown in \fig{volumeint}. Normalizing by the thickness in the out-of-plane direction, $B$, yields an energy release rate per unit length that is constant along the straight crack front:
\begin{equation}
J = \frac{1}{B} \int_{V_0} \left[ P_{jK} \frac{\partial u_j}{\partial X_1} - W \delta_{1K} \right] \frac{\partial q}{\partial X_K}  dV_0.
\label{eq:jdomain}
\end{equation}
Here, $V_0$ is the integration volume in the reference configuration, and $q$ retains the same boundary values defined above while remaining constant in the thickness direction to maintain plane-strain consistency.

The Griffith criterion for fracture is
\begin{equation}
J = G_{\rm c},
\label{eq:Jfraccrit}
\end{equation}
where the material property $G_{\rm c}$ is the critical energy release rate, which for ideally brittle materials (in which all the released energy goes to creating new crack surfaces) equals twice the surface energy, i.e.,\ $G_{\rm c}=2\gamma_{\rm s}$. For most real materials, fracture is accompanied by significant dissipation, and $G_{\rm c} \gg 2\gamma_{\rm s}$.

In the small strain limit where the LEFM framework applies, the stress field in the vicinity of the crack tip exhibits a universal form characterized by the stress intensity factor $K$. For isotropic materials, the energy release rate $G$ is uniquely related to $K$ by
\begin{equation}
G = \frac{K^2}{E'},
\label{eq:gk}
\end{equation}
where $E' = E$ for plane stress conditions, and $E' = E/(1-\nu^2)$ for plane strain, with $E$ being Young's modulus and $\nu$ Poisson's ratio. For the anisotropic materials considered in this work, the relation between $G$ and $K$ follows the complex-variable formulation of Sih et al.~\cite{sih}. The fracture criterion then takes the form
\begin{equation}
K = K_{\rm c},
\label{eq:Kfraccrit}
\end{equation}
where $K_{\rm c}$ is a material's fracture toughness.

Under finite-strain conditions (nonlinear geometry) and/or material nonlinearity, the stress intensity factor $K$ is not defined and \eqn{eq:Kfraccrit} no longer holds. However, $J$ remains applicable for computing the energy release rate, and the fracture criterion in \eqn{eq:Jfraccrit} continues to hold.

\subsection{Quasicontinuum method}
\label{sec:qc}

\begin{figure}[t] \centering
    \includegraphics[trim = 0mm 0mm 0mm 0mm, clip=true, width=0.7
    \textwidth]{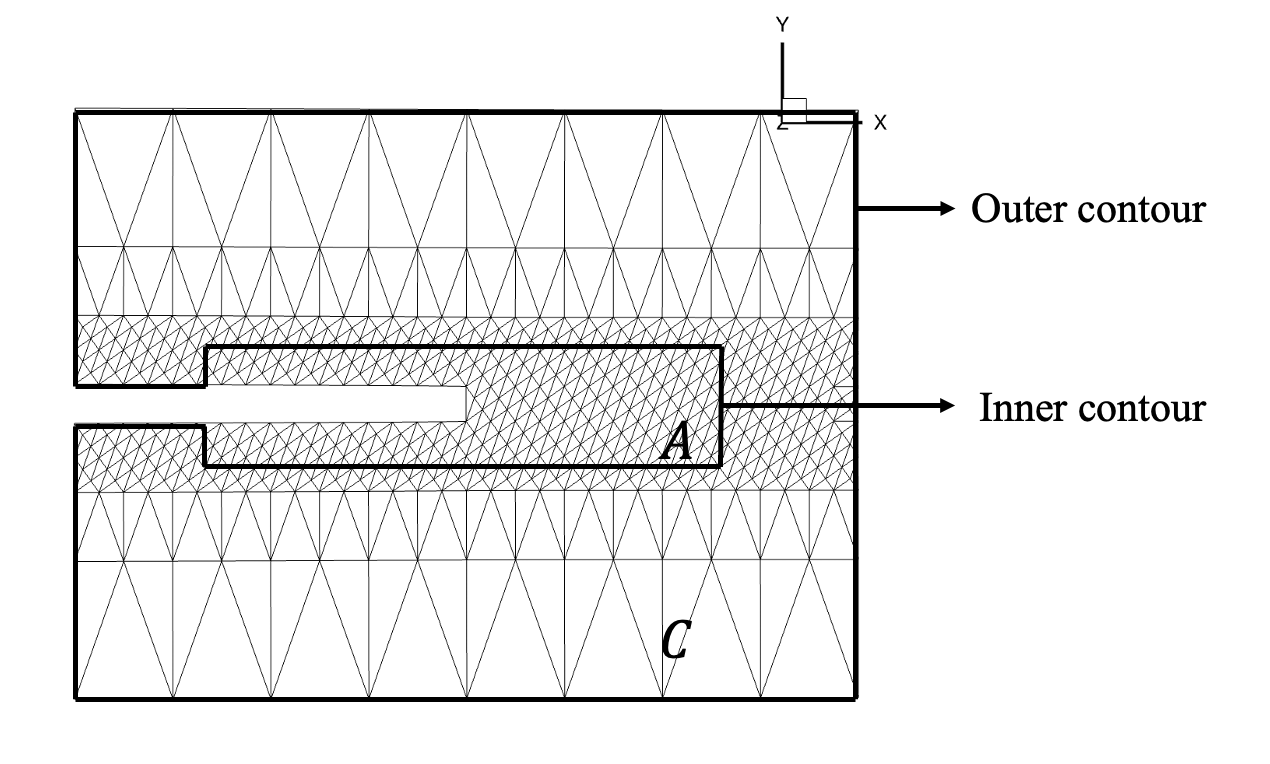}
    \caption{QC model for a crack geometry showing the atomistic region $A$ (inside the inner contour) and the surrounding continuum region $C$. The figure also shows the $J$-integral integration domain bounded by the inner and outer contours.}
    \label{fig:contours}
\end{figure}

The QC method is a multiscale computational framework that seamlessly couples atomistic (nonlocal) regions with continuum (local) regions to model material deformation efficiently across scales. Originally developed by Tadmor et al.~\cite{tadmor:ortiz:1996}, the method enables the simulation of lattice defects, such as dislocations and cracks, within domains that would be computationally challenging for fully atomistic models. The original approach was static, and later extended to equilibrium dynamics at finite temperature where it is called ``Hot QC'' \cite{tadmor:legoll:2013}, with temporal acceleration to reach longer time scales utilizing the hyperdynamics approach added in ``hyper-QC'' \cite{kim:luskin:2014}. QC3D \cite{whalen:2023} is a complete 3D implementation of QC for multilattice materials (i.e.,\ crystals with unit cells containing more than one atom) supporting arbitrary polycrystalline geometries. QC3D uses hybrid parallelization to run efficiently on shared-memory and distributed-memory computing systems, scaling to thousands of compute cores.

In QC, the domain is partitioned into an atomistic region $A$, in which all atoms are retained, and a coarse-grained continuum region $C$ in which a subset of atoms is selected as \emph{representative atoms} or ``repatoms'' for short. Hence, the set of
repatoms includes all atoms in the atomistic region and a small
number of atoms in the continuum region. The repatoms serve as the nodes of an FE mesh, as shown in \fig{fig:contours}. The remaining atoms in the continuum region are not represented explicitly. The displacement field within each element $e$ is linearly interpolated from the displacements of its repatom nodes, so that the deformation gradient $\bF^e$ is uniform. The energy of the element is then given by the Cauchy--Born rule. A single FE mesh spans the entire domain ($A$ and $C$) with the  mesh fully refined down to atomic resolution in the atomistic region. In the static version of the method, the potential energy $\Pi$ is given by\footnote{This is the most basic QC model. For multilattice crystals, the QC potential energy also accounts for the shifts of the basis atoms in the unit cell \cite{tadmor:smith:1999}. At finite temperature, Hot-QC includes molecular dynamics in the atomistic region and free energy calculations in the continuum region \cite{tadmor:legoll:2013}.} \cite{shenoy:miller:1999}
\begin{equation}
\Pi = \sum_{\alpha\in A} E^\alpha(\br_A; \br_P) + \sum_{e\in C} W_{\rm CB}(\bF^e) V^e
- \sum_{\beta\in(A\cup C)} \left(\bff^{\rm ext,\beta} + \bff^{\rm ghost, \beta}\right)\cdot\br^\beta.
\end{equation}
Here, the first term is the energy of the atomistic region in which $E^\alpha$ is the energy of the atom $\alpha$ in the atomistic region $A$, computed using a suitable interatomic potential, which depends on $\br_A$, the positions of the atoms in the atomistic region, and $\br_P$, the positions of padding atoms lying in the continuum region $C$ close to $A$. The second term is the energy of the continuum region, in which $W_{\rm CB}$ is the Cauchy--Born strain energy density evaluated for each element $e$ in the continuum region, with volume $V^e$, in terms of its deformation gradient $\bF^e$ computed from the displacements of the nodes delimiting it. The Cauchy--Born rule assumes that the crystalline structure underlying an element deforms homogeneously according to the element deformation gradient $\bF^e$, so that the lattice vectors transform as $\mathbf{a}_i = \bF^e \mathbf{A}_i$, where $\mathbf{A}_i$ are the reference lattice vectors. For a simple lattice (one atom per unit cell), the strain energy density is then
\begin{equation}
W_{\rm CB}(\bF^e) = \frac{E_{\rm cell}(\bF^e)}{\Omega_0},
\label{eq:cb}
\end{equation}
where $E_{\rm cell}$ is the energy of a unit cell of an infinite perfect crystal subjected to the deformation $\bF^e$, computed using the same interatomic potential as in the atomistic region, and $\Omega_0$ is the reference unit-cell volume. For multilattice crystals, such as silicon that has a diamond-cubic structure with a two-atom basis, the basis atoms can also shift relative to each other in addition to the homogeneous deformation. These shifts enter $E_{\rm cell}$ as additional degrees of freedom in each element and contribute to the total energy \cite{tadmor:smith:1999}. 
The third term includes the potential of the external forces $\bff^{\rm ext,\beta}$, where $\beta$ is any atom in the atomistic region or any node in the continuum region, and $\bff^{\rm ghost,\beta}$ are dead-load correction terms accounting for nonphysical ``ghost forces'' at the atomistic/continuum interface. Ghost forces arise because atoms near the interface interact with repatoms in the continuum region through two inconsistent energy descriptions; they are defined as the nonzero forces that QC computes for a uniformly deformed lattice, for which the exact forces vanish. They are corrected by computing these forces for a uniformly deformed lattice and applying their negatives as $\bff^{\rm ghost,\beta}$ \cite{shenoy:miller:1999}.

Static equilibrium configurations are obtained by minimizing the potential energy with respect to the atom and node positions, $\br = \{\br^\alpha : \alpha \in A\cup C\}$, starting from an initial guess $\br_{\rm init}$ and subject to boundary conditions:
\begin{equation}
\br_{\rm min}
= \underset{\br\in\cC;\ \br_{\rm init}}{\operatorname{arg\,min}} \,\Pi(\br),
\end{equation}
where $\cC$ is the space of allowed configurations given any displacement constraints. Fracture simulations are performed similarly for geometries containing cracks (as in \fig{fig:contours}). The loading progresses in a quasistatic fashion with the external load incrementally increasing at each load step $i$, and the energy is minimized using the result converged previously as an initial guess:
\begin{equation}
\br^{(i)}
= \underset{\br\in\cC^{(i)};\ \br^{(i-1)}}{\operatorname{arg\,min}} \,\Pi(\br).
\end{equation}
This is repeated until fracture is detected. One option is to apply $K$-field displacement boundary conditions to the edges of a model containing a single crack tip, as described in \sect{sec:cracl}, in which case the loading at fracture provides the fracture toughness $K_{\rm c}$. Quasistatic fracture simulations at finite temperature are possible using a free-energy variant of QC \cite{kim:kavalur:2022}.

QC provides the option of automatic mesh adaptation to allow the atomistic region to grow and accommodate an evolving deformation, e.g., crack growth, nucleation of dislocations that move away from the crack tip, or phase transformations \cite{miller:tadmor:1998, shenoy:miller:1999, hai:tadmor:2003}. Unlike conventional FE methods, in which failure is introduced through phenomenological criteria, QC captures fracture in the atomistic region through direct modeling of bond breaking and the accompanying nanoscale processes. QC is therefore a \emph{predictive} method, with accuracy tied to the fidelity of the interatomic potential. For a review of applications of QC to fracture and other multiscale problems, see \cite{miller:tadmor:2002}.

\section{Implementation of $J$-integral in QC}
\label{sec:jinqc}

\subsection{Theoretical formulation}

The $J$-integral integration domain is taken to be coincident with the continuum region, where the Cauchy--Born rule provides well-defined continuum fields. For example, for the model in \fig{fig:contours}, the integration domain is bounded on the inside by a contour along the atomistic/continuum interface and on the outside by a contour along the outer edge of the model. This placement avoids the atomistic region near the crack tip where discrete lattice effects dominate and continuum field definitions become ambiguous.

We generalize \eqn{eq:jdomain} for an arbitrary crack-growth direction $\hat\bn$ by replacing the scalar field $q(\bX)$ with a vector field oriented along $\hat\bn$,
\begin{equation}
\bq(\bX) = q(\bX) \hat\bn.
\label{eq:qvec}
\end{equation}
Here, as before, $q(\bX)$ varies from unity on the inner contour to zero on the outer contour. With this definition, \eqn{eq:jdomain} takes the form
\begin{equation}
J = \frac{1}{B}\int_{V_0} \left[ P_{jK} \frac{\partial u_j}{\partial X_I} - W \delta_{IK} \right] \frac{\partial q_I}{\partial X_K}  dV_0.
\label{eq:genj}
\end{equation}
Equivalently, this can be expressed in terms of Eshelby's tensor $\bH$ (see \eqn{eq:eshelby}),
\begin{equation}
J = -\frac{1}{B}\int_{V_0} H_{IK} \frac{\partial q_I}{\partial X_K}  dV_0.
\label{eq:jeshel}
\end{equation}

\subsection{Finite element discretization}

To implement this formulation within an FE framework, \eqn{eq:genj} must be recast into a discrete form. QC3D employs linear tetrahedral elements within which the Eshelby tensor is constant. The contribution of each element to the total $J$-integral is then
\begin{equation}
J^{e} = -H_{IK}^{e} \left.\frac{\partial{q_I}}{\partial{X_K}}\right|_e V^e,
\label{eq:jele1}
\end{equation}
where $V^e$ is the volume of element $e$. Substituting \eqn{eq:qvec} gives
\begin{equation}
J^{e} = -H_{IK}^{e} \left.\frac{\partial{q}}{\partial{X_K}}\right|_e \hat{n}_{I} V^e.
\label{eq:jele2}
\end{equation}
Using standard FE interpolation, the gradient of $q$ over an element is computed from its nodal values, $q^\alpha$, and the derivatives of the shape functions, $N_{,K}^\alpha \equiv \partial N^\alpha(\bX)/\partial X_K$, so that
\begin{equation}
J^{e} = -H_{IK}^{e} \left( \sum_{\alpha\in\cN^e} q^\alpha N_{,K}^\alpha \right) \hat{n}_{I} V^e,
\label{eq:jele3}
\end{equation}
where $\cN^e$ is the set of nodes at the corners of element $e$ (four for a tetrahedral element). The nodal values of $q$ are defined by assigning $q^{\alpha} = 1$ to the nodes on the inner contour and $q^{\alpha} = 0$ to those on the outer contour. When interpolated with the element shape functions, any choice of interior nodal values yields a continuous, piecewise-linear field that satisfies the requirements of \sect{sec:jint} and is therefore admissible. A linear interpolation of $q$ between the inner and outer contours, introduced by Li et al.\cite{li:shih:1985} is commonly used (we explore the effect of this choice in \sect{sec:sampling}).

The elemental Eshelby tensor, $\bH^e$, is computed from the mechanical fields:
\begin{equation}
H_{IK}^e = W^e\delta_{IK} - P^e_{jK} \left( \sum_{\beta\in\cN^e} u^\beta_j N^\beta_{,I} \right),
\label{eq:geneshel2}
\end{equation}
where $W^e$ and $\bP^e$ are the strain energy density and first Piola--Kirchhoff stress tensor, respectively, both of which are constant within element $e$. The displacement gradient is interpolated using the nodal displacement vectors, $\bu^\beta$, and the derivatives of the shape functions.

Substituting \eqn{eq:geneshel2} into \eqn{eq:jele3} yields the complete discrete expression for the elemental contribution:
\begin{equation}
J^{e} =  \left [P^e_{jK} \left( \sum_{\beta\in\cN^e} u^\beta_jN^\beta_{,I} \right) -W^e \delta_{IK} \right ] \left( \sum_{\alpha\in\cN^e} q^\alpha N_{,K}^\alpha \right) \hat{n}_{I} V^e.
\label{eq:jele4}
\end{equation}
This expression can be succinctly represented in matrix form, highlighting the tensor contractions involved:
\begin{equation}
J^{e} =  \left [ \left.\frac{d\bu}{d\bX}\right|_e^T \bP^e - W^e \bI\right]: \left.\frac{d\bq}{d\bX}\right|_e V^e
\label{eq:jele5}
\end{equation}

This formulation provides the basis for the computational procedure of $J$-integral calculations within the QC framework and integrates seamlessly with the existing FE discretization of the code. Finally, the energy release rate per unit length of crack front follows from
\begin{equation}
J = \frac{1}{B} \sum_{e\in C} J^{e}.
\label{finalj}
\end{equation}
Where $B$ is the thickness. This is equivalent to the standard 2D plane-strain $J$-integral.

\section{Validation of the $J$-integral implementation}
\label{sec:validate}

To validate the $J$-integral implementation within the QC framework, we employ a standard edge-crack geometry under mode~I loading, with the anisotropic LEFM $K$-field displacement solution (following Sih et al.~\cite{sih}) prescribed on all repatoms. Prescribing this asymptotic singular field at the remote boundaries ensures that the crack tip responds as though embedded in an infinite body under arbitrary far-field loading, making this a universal loading configuration independent of specimen geometry, and establishing a direct analytical reference for the energy release rate through the $G$--$K$ relation---valid precisely when $K$-dominance holds \cite{suresh1998fatigue}, i.e.,\ when there exists a region away from the process zone where the LEFM solution provides a sufficiently good approximation to the crack-tip stress field. To extend the validation beyond this regime to cases in which material and geometric nonlinearity break $K$-dominance and render the $G$--$K$ relation inapplicable, we additionally implement the VCE method as an independent numerical measure of the energy release rate.

To construct a systematic validation scheme, three regimes of increasing complexity are considered in \sects{sec:lefm}, \sectx{sec:nonlinear_validation}, and \sectx{sec:full_qc_validation}. In all three, the same prescribed LEFM anisotropic $K$-field displacements are applied to all repatoms at each load step; the regimes differ in the response of the model interior to this field. In the first regime (\sect{sec:lefm}), the stress and strain energy density are computed from Hooke's law. Since the imposed field is the exact equilibrium solution of a linear elastic boundary-value problem, the $J$-integral and VCE should both equal the analytical LEFM energy release rate $G$ without energy minimization. This regime therefore validates both methods against an independently known exact solution. In the second regime (\sect{sec:nonlinear_validation}), the stress and strain energy density are evaluated using the nonlinear Cauchy--Born constitutive relation. The resulting stress field does not satisfy equilibrium because the prescribed displacement field is the equilibrium solution of the linear elastic problem rather than of the nonlinear Cauchy--Born problem, and no energy minimization is performed. Hence, neither the equality $J=G$ nor the path independence of $J$ holds. Similarly, the VCE finite-difference estimate does not equal to $G$ since the potential energy is not minimized at each crack length. However, the energy release rates computed via $J$ and VCE should be mutually consistent if both are implemented correctly. Thus, this regime constitutes a code verification step, as opposed to a physical validation.\footnote{Both quantities correspond to the derivative of the same (non-equilibrium) energy functional with respect to the same prescribed virtual crack advance; $J$ is computed analytically via the Eshelby tensor, and VCE is computed by finite differences. Therefore their agreement follows from calculus alone and does not require stress equilibrium.} In the third and final regime (\sect{sec:full_qc_validation}), a Cauchy--Born constitutive relation is used as in Regime~2, but energy minimization is enabled subject to the same prescribed LEFM $K$-field boundary condition used in Regimes~1 and 2, allowing the system to reach an equilibrium state. No closed-form nonlinear crack-tip solution exists against which this regime can be validated directly. Instead, the close agreement between $J$ and VCE for this true equilibrium configuration provides strong evidence that both methods correctly capture the actual energy release rate of the fully relaxed, nonlinear multiscale system.

\subsection{Crack geometry and quasistatic loading scheme}
\label{sec:cracl}

A schematic of the simulation geometry is provided in \fig{fig:crackgeom}. The initial crack is introduced by removing a single plane of atoms, resulting in a finite initial opening on the order of a few angstroms (the smallest possible physical crack). The computational mesh is constructed with full atomistic resolution near the crack tip, transitioning to a coarser representation in the surrounding continuum region. All simulations are conducted under mode~I loading using a Stillinger--Weber potential for brittle Si that includes a correction in Balamane et al.~\cite{balmane} and a modification by Hauch et al.~\cite{hauch} that leads to brittle fracture, obtained from OpenKIM \cite{tadmor2011potential,minshi}.\footnote{Note that this potential is not meant to provide a realistic model for silicon fracture. It is selected here for computational efficiency since the purpose of the paper is to develop a numerical method for extracting fracture parameters in QC3D, not to study silicon fracture. The choice of potential does not effect the conclusions regarding the methods developed in this work.} Periodic boundary conditions are applied along the crack front direction ($z$) to enforce plane-strain conditions, effectively modeling a crack of infinite extent in this dimension.

\begin{figure}[t]\centering
    \resizebox{0.4\textwidth}{!}{%
    \begin{tikzpicture}[>=latex]
      \def\W{4.5}   
      \def\H{6.0}   
      \def\A{2.25}   
      \def\D{0.3}  
      \draw [line width=1.4pt]
        (0,0) -- (\W,0) -- (\W,\H) -- (0,\H) --
        (0,{\H/2+\D/2}) -- (\A,{\H/2+\D/2}) --
        (\A,{\H/2-\D/2}) -- (0,{\H/2-\D/2}) -- (0,0);
      \draw [<->] ({\W+0.6},0) -- node [right] {$h$} ({\W+0.6},\H);
      \draw [thin] (\W,0) -- ({\W+0.6},0);
      \draw [thin] (\W,\H) -- ({\W+0.6},\H);
      \draw [<->] (0,-0.5) -- node [below] {$w$} (\W,-0.5);
      \draw [thin] (0,0) -- (0,-0.5);
      \draw [thin] (\W,0) -- (\W,-0.5);
      \draw [<->] (0,{\H/2+\D/2+0.6}) -- node [above] {$a$} (\A,{\H/2+\D/2+0.6});
      \draw [thin] (0,{\H/2+\D/2}) -- (0,{\H/2+\D/2+0.6});
      \draw [thin] (\A,{\H/2+\D/2}) -- (\A,{\H/2+\D/2+0.6});
      \draw [<->] ({\A+0.5},{\H/2-\D/2}) -- node [right] {$d$} ({\A+0.5},{\H/2+\D/2});
      \draw [thin] (\A,{\H/2-\D/2}) -- ({\A+0.5},{\H/2-\D/2});
      \draw [thin] (\A,{\H/2+\D/2}) -- ({\A+0.5},{\H/2+\D/2});
      \draw [->] ({\W-2.0},{\H-1.2}) -- ({\W-1.4},{\H-1.2})
        node [right] {$x [\bar{1} 2 \bar{1}]$};
      \draw [->] ({\W-2.0},{\H-1.2}) -- ({\W-2.0},{\H-0.6})
        node [above] {$y [1 1 1]$};
    \end{tikzpicture}%
    }
    \caption{A schematic of the crack geometry considered, indicating key dimensions.}
    \label{fig:crackgeom}
\end{figure}

To drive the system toward nonlinear deformation and eventual fracture instability, the crack geometry is incrementally displaced. At each load step, the boundary nodes are held fixed while the internal atomic and continuum degrees of freedom are relaxed to their equilibrium positions. A significant challenge in this process is the non-convex energy landscape, which can lead to metastable configurations or convergence issues when load steps are too large or initial states are far from equilibrium. To overcome these challenges and enable robust implementation and validation of the $J$-integral across linear and nonlinear regimes, we have developed a tailored loading and relaxation scheme that balances computational efficiency with numerical stability.

The initial displacement field for all repatoms (atoms in the atomistic region and nodes in the continuum region) is prescribed based on the linear elastic solution for a sharp crack in an anisotropic medium under mode~I loading following the formulation of Sih et al.~\cite{sih}. Let $\bu(\bX, K_\rI)$ denote this displacement field where $\bX$ is the reference position of a material point and $K_\rI$ is the mode~I stress intensity factor.  At the first load step, the initial displacement $\bpsi^{\text{init}}_0(\bX, K_{\rI,0})$ is set equal to the linear elastic solution:
\begin{equation}
\bpsi^{\text{init}}_0(\bX, K_{\rI,0}) = \bu(\bX, K_{\rI,0}),
\label{eq:initial_disp}
\end{equation}
where $K_{\rI,0}$ is chosen sufficiently small to ensure that atomic positions near the crack tip remain close to the elastic prediction. The boundary nodes are held fixed at these displacements while all other representative atoms are relaxed via conjugate-gradient minimization of the total potential energy, with a convergence tolerance of $10^{-8}$~\AA{} on the maximum displacement per iteration. The relaxed displacement field is denoted $\bpsi_0(\bX, K_{\rI,0})$.

In subsequent load steps, the initial guess is constructed incrementally rather than being fully re-initialized from the linear elastic solution. Specifically, the displacement increment corresponding to an increase $\Delta K_\rI$ in the stress intensity factor is added to the previously relaxed configuration:
\begin{equation}
\bpsi^{\text{init}}_n(\bX, K_{\rI,n}) = \bpsi_{n-1}(\bX, K_{\rI,n-1}) + \bu(\bX, \Delta K_\rI),
\label{eq:incremental_disp}
\end{equation}
where $K_{\rI,n} = K_{\rI,0} + n \Delta K_\rI$, with $K_{\rI,0} = \Delta K_\rI = 0.05$~eV/\AA$^{5/2}$. This field is then relaxed again to obtain $\bpsi_n(\bX, K_{\rI,n})$, and so on. This incremental approach enables larger load steps than would be possible using only the linear elastic initial guess, while providing a more accurate starting configuration near the crack tip compared to pure linear elasticity. The procedure relies on the assumption that, for small $\Delta K_\rI$, the change in the nonlinear displacement field is also small, so that the initial guess remains near the equilibrium solution.

\begin{figure}[t]
    \centering
    \includegraphics[trim = 0mm 0mm 0mm 0mm, clip=true, width=1.0\textwidth]{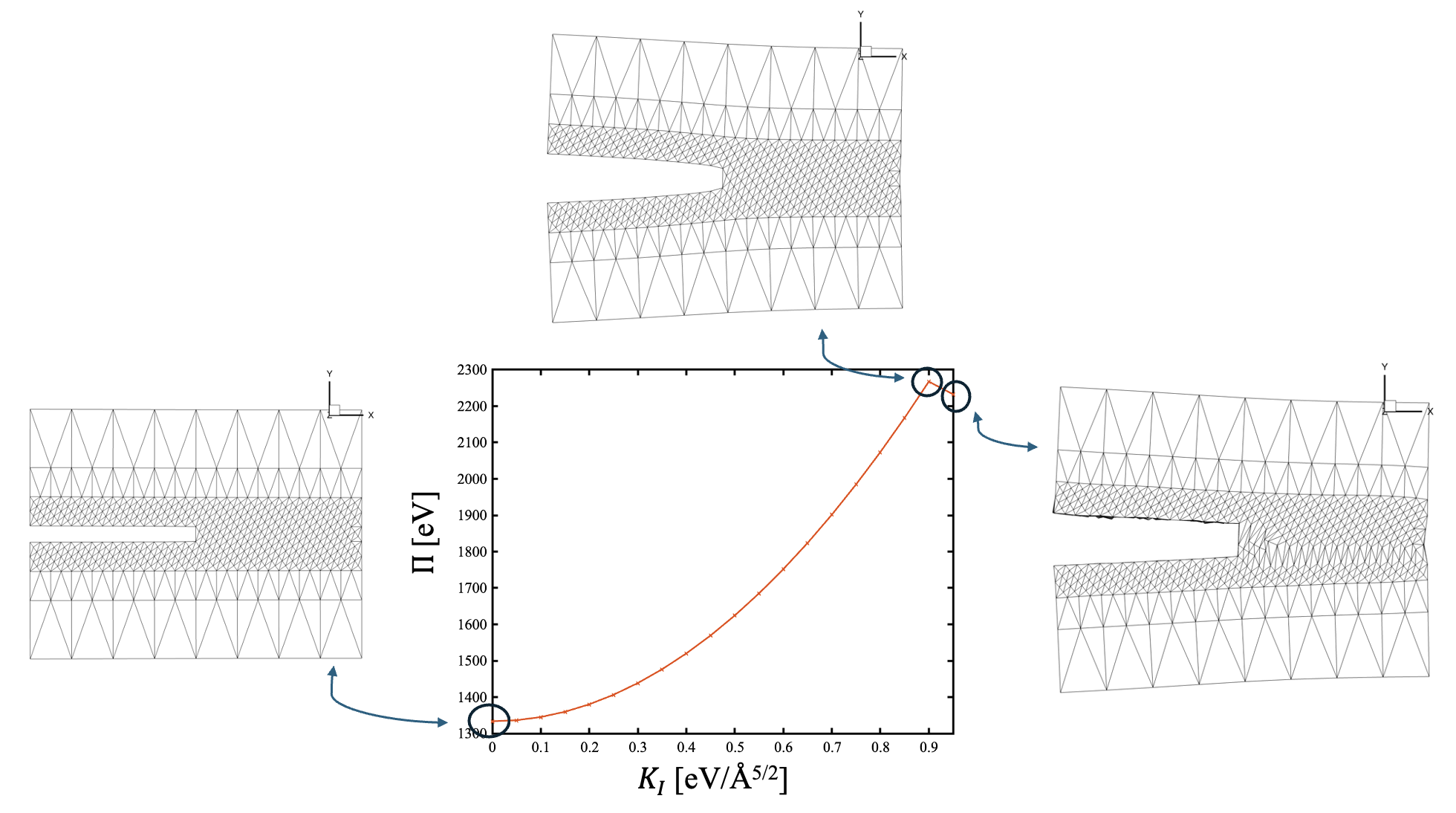}
    \caption{Potential energy $\Pi$ versus the mode~I stress intensity factor $K_\rI$ for silicon modeled using the modified Stillinger--Weber potential. The curve increases monotonically until the energy drops at the point of fracture instability. Snapshots of the deformed QC3D configuration along the loading path are shown.}
    \label{fig:strainenergy}
\end{figure}

The onset of fracture is automatically detected by monitoring discontinuities in the total potential energy $\Pi$. Under displacement-controlled quasistatic loading, $\Pi$ is equal to the strain energy and increases monotonically until cleavage, at which point a sudden drop signals the loss of load-bearing capacity. The detection algorithm (\alg{alg:cleavage_convergence}) extrapolates $\Pi$ linearly from the two preceding load steps and flags cleavage when the deviation exceeds $f$ times the previous energy, where $f$ is a prescribed tolerance factor ($f=0.1$ in this work), at which point the simulation is terminated. \fig{fig:strainenergy} shows a plot of $\Pi$ versus $K_\rI$ up to the fracture instability for silicon modeled using the modified Stillinger--Weber potential, together with snapshots of deformed QC3D configurations along the loading path. The last snapshot shows that the crack has propagated along the length of the model---note that the mesh in the atomistic region contains vacuum elements; therefore, the vertically stretched elements to the right of the crack tip connect atoms that have been separated beyond their range of cohesion.

\begin{algorithm}[t]
\caption{Crack cleavage detection via strain energy monitoring}
\begin{algorithmic}[1]
\State \textbf{Input:} Current strain energy $\Pi_{\text{curr}}$, tolerance factor $f$
\State \textbf{Output:} Cleavage status (True/False)

\State $ \text{visit\_count} \gets \text{visit\_count} + 1$
\State $ \Pi_{k-2} \gets \Pi_{k-1}$
\State $ \Pi_{k-1} \gets \Pi_{k}$
\State $ \Pi_{k} \gets \Pi_{\text{curr}}$ \Comment{Update energy history}

\If{$\text{visit\_count} > 2$} \Comment{Require at least three data points}
    \State $\Pi_{k}^{\text{pred}} \gets 2\Pi_{k-1} - \Pi_{k-2}$ \Comment{Linear extrapolation}
    \State $\delta \gets |\Pi_{k} - \Pi_{k}^{\text{pred}}|$ \Comment{Deviation from prediction}
    \If{$\delta > f\, |\Pi_{k-1}|$}
        \State \Return \texttt{True} \Comment{Cleavage detected; terminate simulation}
    \Else
        \State \Return \texttt{False} \Comment{Continue loading}
    \EndIf
\Else
    \State \text{OutputMessage}(``Insufficient data points for cleavage test'')
    \State \Return \texttt{False}
\EndIf
\end{algorithmic}
\label{alg:cleavage_convergence}
\end{algorithm}

\subsection{Virtual crack extension method}
\label{sec:vce}

The connection between the domain $J$-integral formulation and the VCE method provides an important physical insight into the nature of the $q$-field introduced in \eqn{eq:jdomain2D}: specifically, the $q$-field may be interpreted as a virtual crack extension field that weights the material perturbation associated with an infinitesimal crack advance. Parks \cite{parks:1974} and Hellen \cite{hellen:1975} independently developed the VCE method as an FE technique to compute energy release rates. deLorenzi \cite{delorenzi:1982} later derived a generalized analytical expression for the energy release rate valid for elastic materials and materials obeying the deformation theory of plasticity, showing that the VCE formulation reduces to the domain $J$-integral with the nodal shifts represented by the $q$-field. This theoretical connection motivated our use of the VCE method to validate the $J$-integral results at finite deformation where LEFM is not applicable.

The VCE method computes the energy release rate, $-d\Pi/dA$, by considering the energy difference between models having a reference crack and a virtually extended crack of slightly greater length. In conventional FE analysis this can be achieved by perturbing the crack tip and surrounding nodes along the propagation direction within a single mesh and then computing the energy release rate from the energy difference between these configurations under identical loading conditions. In the QC method, however, the essential procedure of energy minimization (relaxation) prevents the evaluation of both configurations within a single simulation. Performing the VCE calculation without relaxation would reduce QC to a conventional FE method with an atomistically resolved mesh, thereby nullifying its multiscale character.

\begin{figure}[t]\centering
    \includegraphics[trim = 0mm 0mm 0mm 0mm, clip=true, width=0.7
    \textwidth]{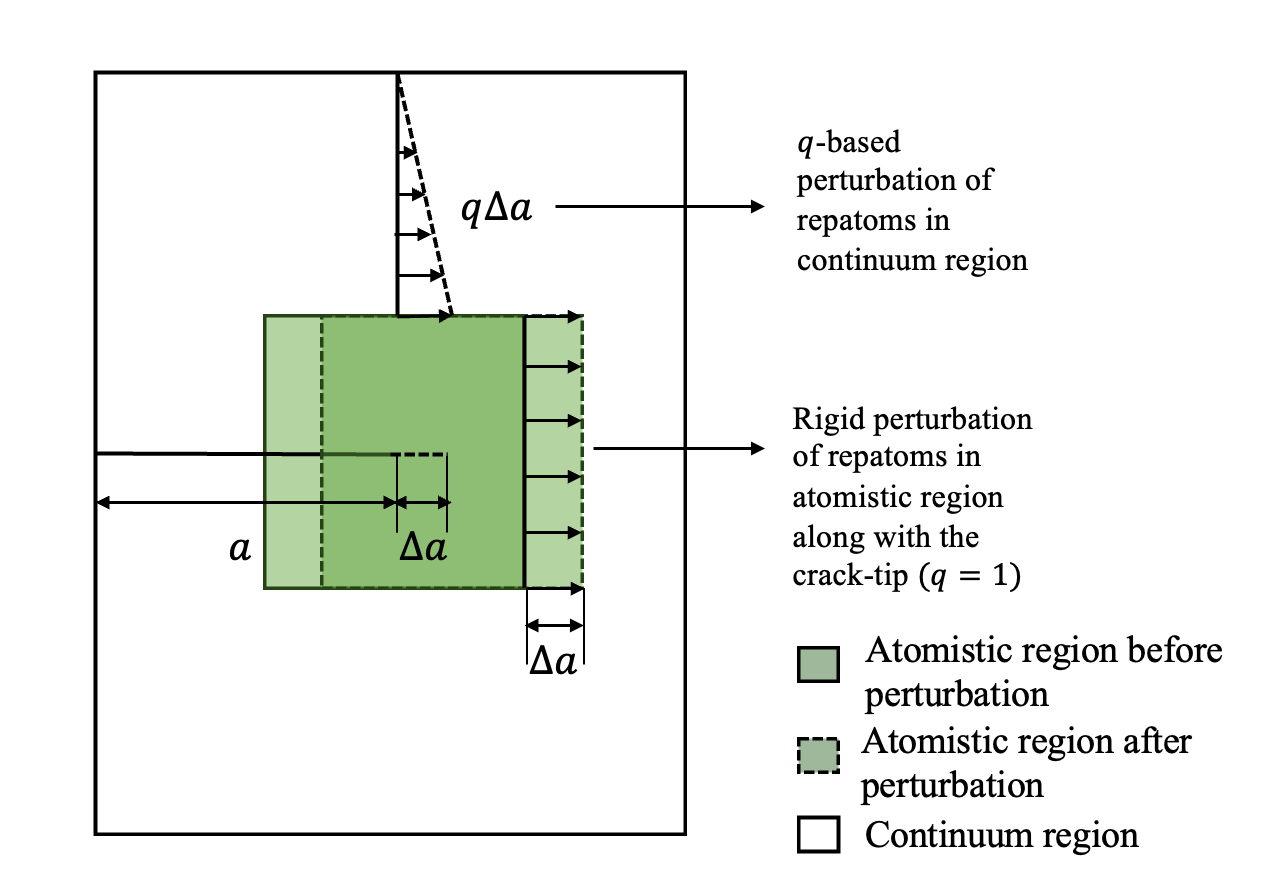}
    \caption{A schematic of the repatom perturbation during virtual crack extension.}
    \label{fig:vce}
\end{figure}

Consequently, our implementation requires two separate, fully relaxed QC simulations: one for the reference crack length and another for the virtually extended crack. The energy release rate is then computed in a post-processing step from the difference in total strain energy between these equilibrated configurations at corresponding load steps, thereby properly accounting for atomic-scale relaxations. \fig{fig:vce} illustrates the repatom perturbation scheme. The crack tip and the repatoms in the atomistic region surrounding it are rigidly translated by an amount $\Delta a$ in the direction of crack growth. For repatoms in the continuum region, the perturbation is governed by the $q$-field, in which each node is shifted by $q^\alpha \Delta a$. Here, $q^\alpha$ are the nodal values of the linearly interpolated $q$-field used in the $J$-integral computation. This strategy, which follows the recommendation of deLorenzi~\cite{delorenzi:1982}, ensures consistency between the VCE and $J$-integral methods and allows a direct comparison of their results.

We emphasize that the virtual crack extension is a geometric modification, not a kinematic perturbation. The nodal shifts define a new reference geometry with an extended crack length; they are not additional displacements applied to the original configuration. Both the reference and extended configurations are subjected to the same $\Delta K$ and the same anisotropic LEFM displacement solution, but the resulting displacement fields differ for two reasons: the field is evaluated relative to different crack-tip positions, and the nodal positions themselves have shifted according to the $q$-field (see \fig{fig:vce_kfield}). This geometric distinction ensures that the energy difference accurately reflects the energy release rate associated with crack advance. The implementation logic for VCE in a non-relaxing context is shown in \alg{alg:vceimplementation}. In our QC3D application, this logic is applied across two separate, fully relaxed simulations at each load step to compute $-d\Pi/dA$.

\begin{figure}[t]\centering
    \begin{subfigure}[b]{0.47\textwidth}
        \centering
        \resizebox{\textwidth}{!}{%
        \begin{tikzpicture}[>=latex]
          \footnotesize
          \useasboundingbox (0,-0.5) rectangle (5.5,4.5);
          \def\W{5.0}
          \def\H{4.0}
          \def\A{2.0}
          \def\D{0.1}
          \draw [line width=1.0pt]
            (0,0) -- (\W,0) -- (\W,\H) -- (0,\H) --
            (0,{\H/2+\D/2}) -- (\A,{\H/2+\D/2}) --
            (\A,{\H/2-\D/2}) -- (0,{\H/2-\D/2}) -- (0,0);
          \fill (\A,{\H/2}) circle (2.5pt);
          \node [below] at (\A,{\H/2-0.15}) {$\bx_{\text{tip}}$};
          \fill [blue] (3.6,3.0) circle (2pt);
          
          \draw [->, thick, magenta] (\A,{\H/2}) -- (3.6,3.0)
            node [midway,right, xshift=2pt] {$\bX-\bx_{\text{tip}}$};
          \draw [->, thick, blue] (3.6,3.0) -- (3.85,3.3);
          \node [right] at (2.0,3.5) {$\Delta\bu(\bX-\bx_{\text{tip}},\Delta K)$};
        \end{tikzpicture}%
        }
        \caption{Reference configuration}
        \label{fig:vce_ref}
    \end{subfigure}
    \hfill
    \begin{subfigure}[b]{0.47\textwidth}
        \centering
        \resizebox{\textwidth}{!}{%
        \begin{tikzpicture}[>=latex]
          \footnotesize
          \useasboundingbox (0,-0.5) rectangle (5.5,4.5);
          \def\W{5.0}
          \def\H{4.0}
          \def\Ap{2.5}
          \def\D{0.1}
          \draw [line width=1.0pt]
            (0,0) -- (\W,0) -- (\W,\H) -- (0,\H) --
            (0,{\H/2+\D/2}) -- (\Ap,{\H/2+\D/2}) --
            (\Ap,{\H/2-\D/2}) -- (0,{\H/2-\D/2}) -- (0,0);
          \fill (\Ap,{\H/2}) circle (2.5pt);
          \node [below] at (\Ap,{\H/2-0.15}) {$\bx_{\text{tip}}+\Delta a$$\hat\bn$};
          \fill [blue] (3.7,3.0) circle (2pt);
          
          \draw [->, thick, magenta] (\Ap,{\H/2}) -- (3.7,3.0)
            node [midway, xshift=2pt] {$\bX+\bq\Delta a-(\bx_{\text{tip}}+\Delta a\hat\bn)$};
          \draw [->, thick, blue] (3.7,3.0) -- (4.2,3.4);
          \node [right] at (-0.09,3.6) {$\Delta\bu(\bX+\bq\Delta a-(\bx_{\text{tip}}+\Delta a\hat\bn),\Delta K)$};
        \end{tikzpicture}%
        }
        \caption{Virtually extended configuration}
        \label{fig:vce_pert}
    \end{subfigure}
    \caption{Schematic of the (a) reference, and (b) virtually extended configurations used in VCE. The difference in $\Delta \bu$ fields between (a) and (b) emerges from the difference in positions vectors of repatoms with respect to crack-tip. The relation between $\bq$ and $\hat\bn$ follows \eqn{eq:qvec}.}
    \label{fig:vce_kfield}
\end{figure}

\alg{alg:vceimplementation} summarizes this procedure in pseudocode. The function \textsc{ComputeEnergy} evaluates the total strain energy of the model at a given crack-tip position: the displacement $\bu^{(i)}$ of each repatom $i$ (in the range 1 to $n_{\text{rep}}$, spanning both the atomistic and continuum regions) is incremented by the anisotropic linear-elastic $K$-field displacement $\Delta\bu$ due to an increment $\Delta K$ in the SIF. This is implemented in the \texttt{lefm\_aniso\_disp} function, which returns the mode-I anisotropic LEFM displacement field due to Sih et al. \cite{sih} for a given repatom's position relative to the current crack tip. The total strain energy $\Pi$ is then accumulated over all $n_{\text{elements}}$ elements, each contributing its energy density times its reference volume: for continuum elements this density is the Cauchy--Born energy density from the element deformation gradient, while for atomistic elements it is derived from the interatomic potential energy of the surrounding repatoms, each weighted by its Voronoi volume share of the element. Mixed elements\footnote{Elements having both local and nonlocal repatoms as nodes.} accumulate both contributions, so that $\Pi$ correctly represents the strain energy of the full atomistic-to-continuum model. The algorithm calls \textsc{ComputeEnergy} twice to obtain $\Pi_1$ (reference) and $\Pi_2$ (perturbed). Between the two calls, the geometry is perturbed as $X_{1,\text{pert}}^{(i)} = X_{1}^{(i)} + q^{(i)} \Delta a$; atomistic nodes carry $q^{(i)}=1$ and translate rigidly by $\Delta a$, while continuum nodes carry $0 \le q^{(i)} \le 1$ according to the linearly interpolated $q$-field used in the $J$-integral computation with $q^{(i)}=1$ on the inner contour and $q^{(i)}=0$ on the outer contour.\footnote{Reversing this convention, i.e., $q^{(i)}=0$ on the inner contour and $q^{(i)}=1$ on the outer contour, reverses the sign of the computed $J$-integral while leaving its magnitude unchanged, since the domain integral is linear in $\nabla q$. The original sign is recovered if the crack-growth direction $\hat\bn$ is reversed simultaneously. We adopt the convention stated above, which preserves the physical interpretation of $q$ as a virtual crack extension field.} The crack tip is then advanced by $\Delta a$ before the displacement field and $\Pi_2$ are recomputed. Finally, $G$ is obtained from the forward-difference estimate, $G=-(\Pi_2-\Pi_1)/(B\Delta a)$. In this work, $\Delta a = 10^{-5}$~\AA{} is used.

\begin{algorithm}[t!]
\caption{Virtual crack extension implementation in QC3D for energy release rate computation}
\begin{algorithmic}[1]
\State \textbf{Input:} Reference repatom positions $\bX_{\text{ref}}$, displacements $\bu_{\text{init}}$, crack tip position $\bx_{\text{tip}}$, $\Delta K$, extension length $\Delta a$, thickness $B$
\State \textbf{Output:} Energy release rate $G = -d\Pi/dA$
\Statex
\Function{ComputeEnergy}{$\bX$, $\bu$, $\bx_{\text{tip}}$}
    \For{$i = 1$ to $n_{\text{rep}}$}
        \State $\Delta\bu \gets \text{lefm\_aniso\_disp}(\bX^{(i)} - \bx_{\text{tip}}, \Delta K)$
        \State $\bu^{(i)} \gets \bu^{(i)} + \Delta\bu$ \Comment{Apply mode-I $K$-field displacement}
    \EndFor
    \State $\Pi \gets 0$
    \For{$e = 1$ to $n_{\text{elements}}$}
        \State $\Pi \gets \Pi + \Pi_e$ \Comment{Accumulate elemental strain energy}
    \EndFor
    \State \Return $\Pi$, $\bu$
\EndFunction
\Statex
\State \textbf{Step 1: Energy at original crack-tip position}
\State $\Pi_1, \bu \gets \Call{ComputeEnergy}{\bX_{\text{ref}}, \bu_{\text{init}}, \bx_{\text{tip}}}$
\Statex
\State \textbf{Step 2: Generate perturbed configuration}
\State $\bu \gets \bu_{\text{init}}$ \Comment{Reset displacements}
\State $\bX_{\text{pert}} \gets \bX_{\text{ref}}$
\For{$i = 1$ to $n_{\text{rep}}$}
    \State $\bX_{\text{pert}}^{(1,i)} \gets \bX_{\text{pert}}^{(1,i)} + q^{(i)} \Delta a$ \Comment{Perturbation governed by the $q$-field in $x$-direction}
\EndFor
\Statex
\State \textbf{Step 3: Energy at advanced crack-tip position}
\State $\bx_{\text{tip}}^{(1)} \gets \bx_{\text{tip}}^{(1)} + \Delta a$  \Comment{Advance crack-tip in $x$-direction}
\State $\Pi_2, \bu \gets \Call{ComputeEnergy}{\bX_{\text{pert}}, \bu, \bx_{\text{tip}}}$
\Statex
\State \textbf{Step 4: Energy release rate}
\State $G \gets -(\Pi_2 - \Pi_1) / (B\Delta a)$
\Comment{First-order forward-difference estimate}
\State \Return $G$
\end{algorithmic}
\label{alg:vceimplementation}
\end{algorithm}

\begin{figure}[t]\centering
     \begin{subfigure}[b]{0.4\textwidth}
         \centering
         \includegraphics[trim = 0mm 0mm 0mm 0mm, clip=true,width=\textwidth]{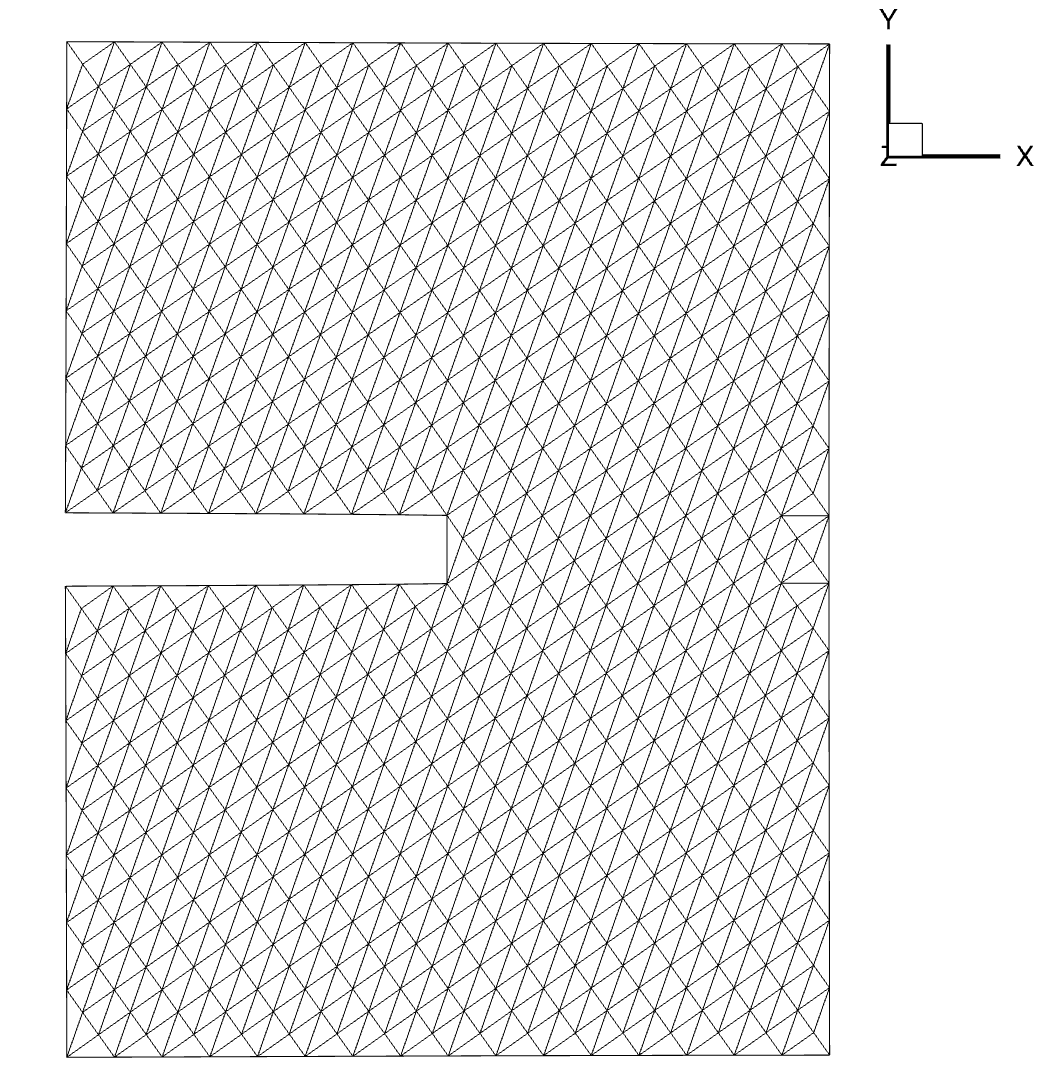}
         \caption{}
         \label{fig:refinedmesh}
     \end{subfigure}
          \begin{subfigure}[b]{0.4\textwidth}
         \centering
         \includegraphics[trim = 0mm 0mm 0mm 0mm, clip=true,width=\textwidth]{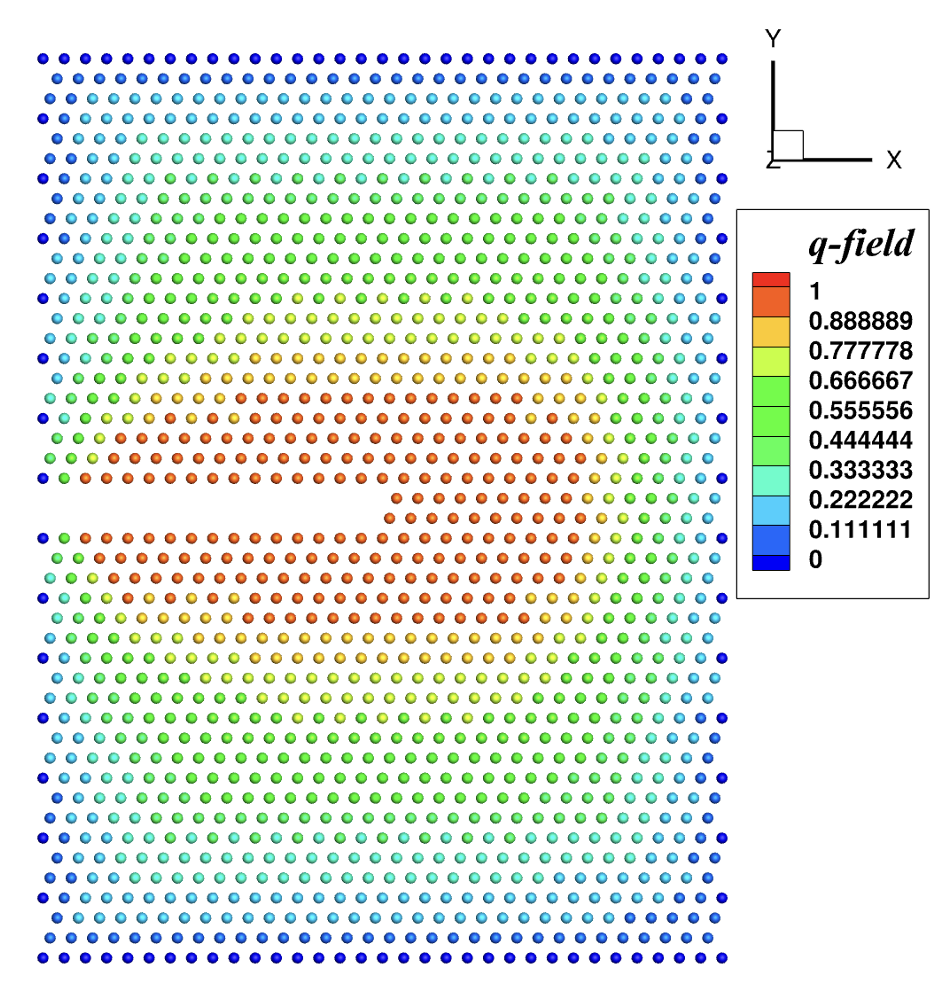}
         \caption{}
         \label{fig:qfieldline}
     \end{subfigure}
     \caption{Model configuration used in the validation studies: (a) fully refined mesh; (b) linearly interpolated $q$-field over the mesh.}
     \label{fig:meshconfig}
\end{figure}

To ensure the highest possible accuracy for our benchmark calculations, a fully-refined mesh is employed for all subsequent computations as shown in \fig{fig:refinedmesh}. In this configuration, the continuum region of the QC3D model that is typically coarsened is instead meshed at atomic resolution. The results for both the $J$-integral and the VCE energy release rate presented in the following sections are computed using this fully-refined model.
In addition, \fig{fig:qfieldline} shows the linearly interpolated $q$-field in the integration domain. The $q$-values of the nodes in the atomistic region do not enter the $J$-integral calculations. However, they are assigned a value of 1 to support the VCE method.

\subsection{Validation against the linear elastic solution}
\label{sec:lefm}

We begin by validating our numerical framework against the established solutions of LEFM. This provides a critical benchmark to verify our implementations of the $J$-integral and VCE methods before proceeding to more complex, nonlinear simulations. To ensure strict adherence to linear elastic conditions, we employ a generalized Hooke's law constitutive model in the continuum region---utilizing the infinitesimal strain tensor and Cauchy stress within our finite-deformation framework---and suppress atomic relaxation. This approach eliminates both material and geometric nonlinearities, creating an idealized linear elastic system in which the anisotropic analytical energy release rate $G = f(K_\rI)$ provides the exact reference solution.

\begin{figure}[t]
    \centering
    \includegraphics[trim = 0mm 0mm 0mm 0mm, clip=true, width=0.7
    \textwidth]{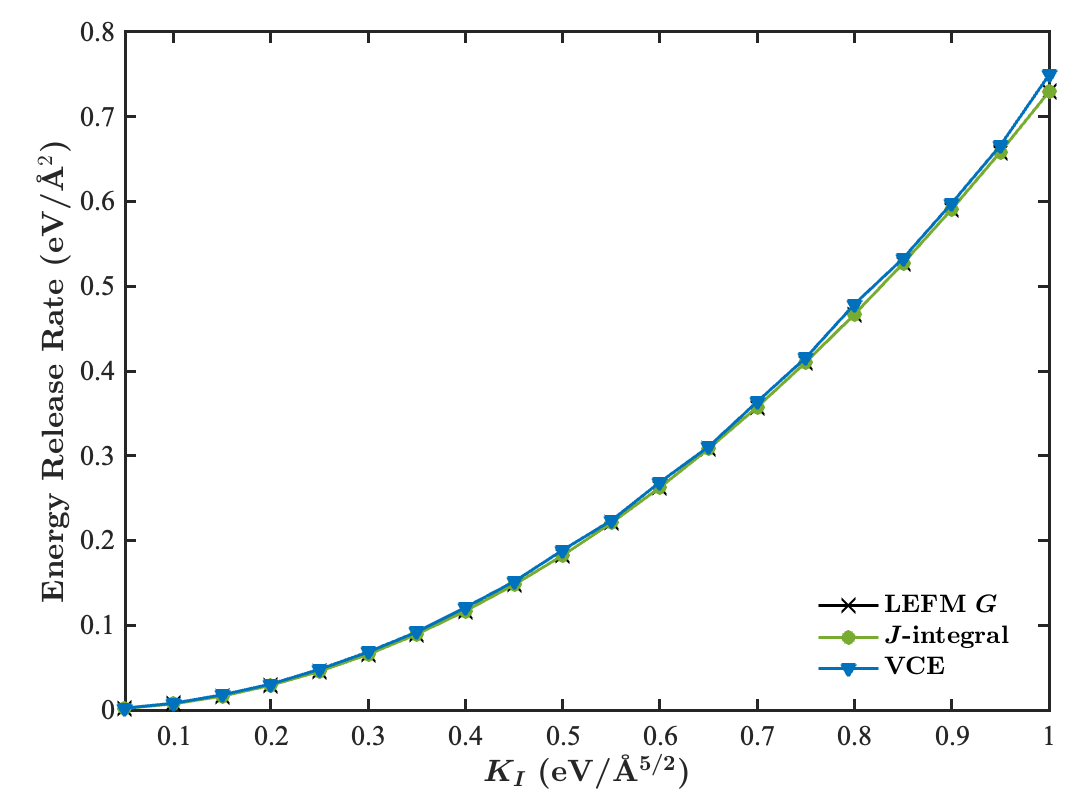}
    \caption{Comparison of energy release rates computed from the analytical LEFM solution, the $J$-integral, and the VCE method in the small strain limit.}
    \label{fig:linelasvalid}
\end{figure}

\fig{fig:linelasvalid} shows the energy release rates computed via the analytical LEFM solution, the domain-form $J$-integral, and the VCE method as a function of the applied mode~I stress intensity factor, $K_\rI$. The results from all three methods are nearly indistinguishable and exhibit the expected quadratic dependence on $K_\rI$. The employed constitutive model sets the first Piola-Kirchhoff stress tensor $\bP$ used in the $J$-integral calculation equal to the Cauchy stress $\boldsymbol{\sigma}$, maintaining direct consistency with classical LEFM theory. This close agreement confirms the numerical fidelity of our implementations and verifies that both the $J$-integral and VCE methods correctly capture the energy release rate in the linear elastic regime.\footnote{The small discrepancy at the last point is due to floating point precision errors in the VCE calculation, which involves division of a finite-difference quantity $\Pi_2-\Pi_1$ (a small difference between two large numbers that scale with $K$) by $\Delta a$.}

\subsection{Verification under material and geometric nonlinearity}
\label{sec:nonlinear_validation}

\begin{figure}[t]
    \centering
    \includegraphics[trim = 0mm 0mm 0mm 0mm, clip=true, width=0.7
    \textwidth]{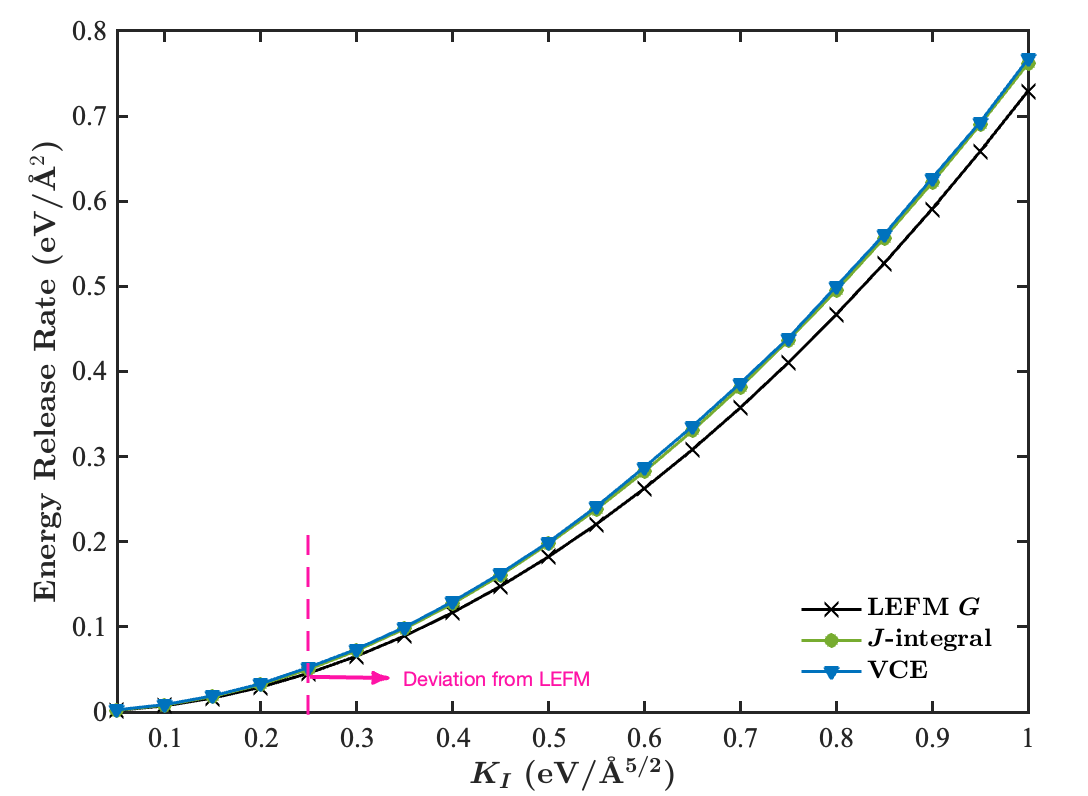}
    \caption{Comparison of energy release rates computed from the analytical LEFM solution, the $J$-integral, and the VCE method with material and geometric nonlinearity.}
    \label{fig:matnonlin}
\end{figure}

Having established the numerical fidelity of our implementation within the linear elastic regime, we now proceed to verify the framework in the presence of material and geometric nonlinearity. To this end, the Cauchy--Born model, derived from the interatomic potential, is reinstated in the continuum region, introducing both the nonlinear constitutive response (material nonlinearity) and finite deformation kinematics (geometric nonlinearity) inherent to this model. Atomic relaxation remains suppressed, so all repatoms follow the prescribed anisotropic $K$-field displacement exactly. This isolates the continuum model's nonlinear response from the atomic-scale deviations introduced by relaxation, providing a critical test of the $J$-integral and VCE methods beyond the infinitesimal strain limit.

\fig{fig:matnonlin} shows the energy release rate computed via the $J$-integral, the VCE, and the classical LEFM solution as a function of the applied far-field load, characterized by the linear elastic $K_\rI$ field used as the boundary condition. At lower load levels, where deformations remain small, the $J$-integral and VCE results coincide with the LEFM solution. However, as the load increases beyond $K_\rI \approx 0.25$~eV/\AA$^{5/2}$, where finite-deformation effects become more pronounced, a substantial deviation from the LEFM prediction is observed. Importantly, the $J$-integral and VCE results remain in excellent agreement with each other throughout the entire deformation path.

The divergence from LEFM reflects the geometric and material nonlinearity inherent in this finite-strain Cauchy--Born model. As discussed in \sect{sec:validate}, the agreement between $J$ and VCE here is expected on calculus grounds alone and does not by itself confirm that either quantity equals the true energy release rate. Rather, it verifies that both methods work correctly and consistently under the nonlinear constitutive relation, establishing confidence in this implementation ahead of the final validation in \sect{sec:full_qc_validation}, where energy minimization is applied to obtain an equilibrium nonlinear solution.

\subsection{Validation with atomic relaxation}
\label{sec:full_qc_validation}

\begin{figure}[t]
    \centering
    \includegraphics[trim = 0mm 0mm 0mm 0mm, clip=true, width=0.7
    \textwidth]{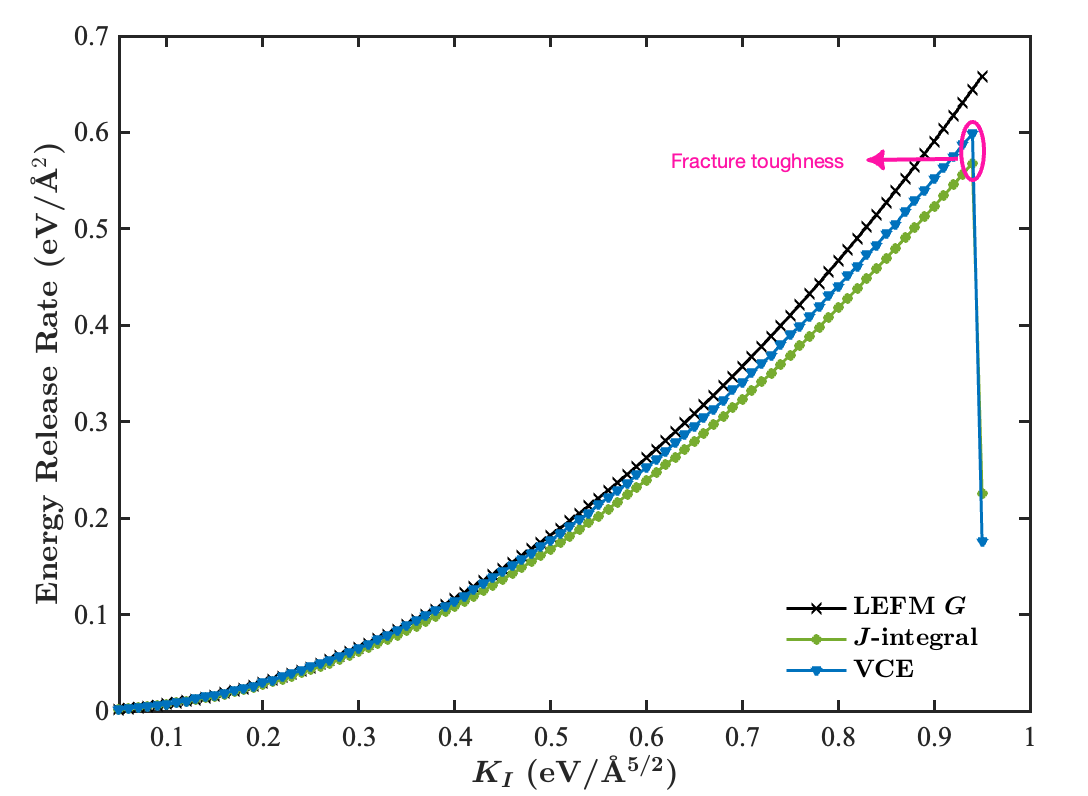}
    \caption{Comparison of energy release rates computed from the analytical LEFM solution, the $J$-integral, and the VCE method with relaxation.}
    \label{fig:fullnonlin}
\end{figure}

The final and most comprehensive validation incorporates the complete physics of the QC method: the geometric and material nonlinearity already present in \sect{sec:nonlinear_validation}, together with relaxation at each load step allowing atoms and nodes to deviate from the continuum solution to adopt their equilibrium positions.
This is the intended operating regime of the model, in which complex nonlinear atomistic relaxation processes not captured by classical continuum theories can occur. Unlike the linear elastic and nonlinear validations above, where the absence of atomic relaxation allowed both the reference and perturbed energies to be evaluated directly on a single mesh via prescribed analytical displacements (as in \alg{alg:vceimplementation}), the present fully relaxed validation requires the full two-simulation VCE procedure described in \sect{sec:vce}: two independently relaxed QC3D simulations, at the reference and virtually extended crack lengths, are used to obtain $\Pi_1$ and $\Pi_2$ at each load step.

\begin{figure}[t]
    \centering
    \includegraphics[trim = 0mm 0mm 0mm 0mm, clip=true, width=0.3\textwidth]{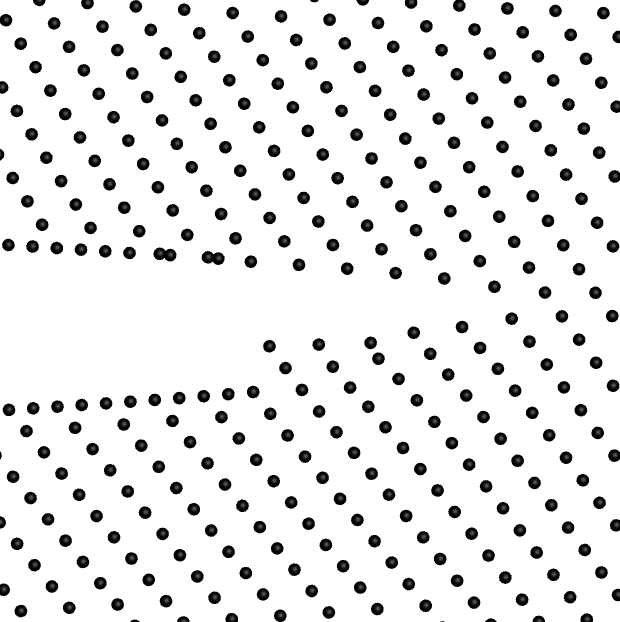}
    \caption{Snapshot of crack propagation in the atomistic region at an applied load of $K_\rI = 0.95$~eV/\AA$^{5/2}$, which corresponds to the sudden drop in energy release rate shown in \fig{fig:fullnonlin}}.
    \label{fig:crackprop}
\end{figure}

\fig{fig:fullnonlin} shows the energy release rates computed from the analytical LEFM solution, the $J$-integral, and the VCE method. At small applied loads, all three agree, confirming that the linear elastic far-field response is correctly transmitted to the crack-tip region. However, as finite deformation develops, the numerical results from the $J$-integral and VCE systematically deviate from the LEFM prediction. This deviation reflects the combined effects of geometric and material nonlinearity, together with atomic-scale relaxation effects, including surface relaxation and possible lattice trapping at the atomistic crack tip, which alter the local driving force for fracture.

Importantly, the $J$-integral and VCE methods maintain good agreement with each other throughout the loading history, with a maximum observed discrepancy of approximately 3\% at the final stable load step immediately preceding fracture instability, the point identified as the fracture toughness.\footnote{The 3\% discrepancy can be attributed to the inherent challenges of the finite-difference approximation in VCE for a nonlinear energy landscape.}
As discussed in \sect{sec:validate}, no closed-form nonlinear crack-tip solution exists against which this regime can be validated directly. Instead the close correspondence observed between the $J$ and VCE calculations provides strong evidence that both methods correctly capture the energy release rate of the equilibrated, fully nonlinear system, even as the process zone undergoes complex atomic-scale rearrangements. At the final load step, both the $J$-integral and VCE values drop suddenly as the crack propagates, as shown in \fig{fig:crackprop}. 

Taken together, the three regimes described in \sects{sec:lefm}, \sectx{sec:nonlinear_validation}, and \sectx{sec:full_qc_validation} confirm that the proposed $J$-integral implementation agrees with an independent linear elastic solution (Regime 1), internally consistent once nonlinearity is introduced (Regime 2), and in close agreement with an independently constructed method on the fully equilibrated, nonlinear multiscale system (Regime 3)---establishing confidence in its use as a measure of the crack driving force across the full range of physics captured by QC.

\section{Statistical assessment of integration-domain quality using $q$-field uncertainty}
\label{sec:sampling}

The $q$-field in the domain form of the $J$-integral defines a virtual crack extension within an integration domain bounded by the inner contour ($q=1$) and the outer contour ($q=0$). Theoretically $J$ is the same for any admissible $q$-field (\sect{sec:jint}), however this no longer holds for discretized numerical methods. As shown in \app{app:contour-to-domain}, the domain form of the $J$-integral evaluated with discrete fields contains a residual $R_h(q)$ that arises from the discretization and vanishes for exact fields. The dependence of $J$ on $q$ results from this residual alone.

Shih et al.~\cite{shih:moran:1986} investigated the influence of the choice of the $q$-field on $J$-integral results by comparing different functional forms, such as ``pyramid'' and ``plateau.'' They showed that near-tip domains yield deviant $J$-values and that smoother $q$-fields can mitigate adverse mesh effects. deLorenzi~\cite{delorenzi:1982} observed that different nodal shift patterns---equivalent to different $q$-fields---should yield the same energy release rate but that numerical discretization introduces scatter, and therefore averaged the results of a small set of nodal perturbations. These studies probed only a few hand-picked $q$-fields and gave limited insight into the quality of the continuum fields in the integration domain.

Since every admissible $q$-field is an equally valid choice, the spread of $J$ over the admissible $q$-fields measures how reproducible a given calculation is with respect to the choice of $q$, i.e., its \emph{precision}. This setting fits naturally within the framework of uncertainty quantification (UQ), in which an input that is not uniquely determined is treated as a random variable and the resulting spread of the quantity of interest (QOI) is used as a measure of uncertainty \cite{wen:tadmor:2020}. Here, the input is the $q$-field and the QOI is $J$. Since the $q$-field is a purely numerical choice, the resulting uncertainty is a numerical uncertainty, and, as for other UQ measures, it characterizes the precision of the result rather than its accuracy. 

We quantify this uncertainty, hereafter referred to as the $q$-field uncertainty or simply the uncertainty, by the standard deviation $\sigma_J$ obtained from Markov chain Monte Carlo (MCMC) sampling of the admissible $q$-fields. Because the $q$-dependence arises from the residual $R_h(q)$ (\eqn{eq:jh_residual}), this uncertainty originates from the discretization error of the fields in the integration domain, which is large when the elements are coarse relative to the length scale over which the fields vary, such as near the crack tip or in overly coarse meshes. $\sigma_J$ thus reflects how well the discrete fields satisfy configurational equilibrium there and therefore provides an intrinsic indicator of the quality of the integration domain and, consequently, of the $J$-integral calculation.

After describing the sampling procedure and verifying its convergence (\sects{sec:sampproc} and \sectx{sec:convdiag}), we examine $\sigma_J$ under two changes that affect the fields in the integration domain: the size and placement of the integration domain (\sect{sec:domainplace}) and the coarsening of the continuum mesh (\sect{sec:meshcoarsening}). Since the procedure operates only on the nodal $q$-values and the elemental contributions $J^e$, it applies to any FE discretization; this is demonstrated on a conventional FE calculation in \sect{sec:fedemo}.

\subsection{Sampling procedure}
\label{sec:sampproc}

\alg{alg:mcmc_qfield} summarizes the procedure. The chain starts from the linearly interpolated field $q_{\text{lin}}$. At each step, one of the $N_{\text{int}}$ interior nodes of the integration domain is selected at random, and its $q$-value is replaced by a value drawn from $\text{Uniform}(0,1)$\footnote{Drawing the new nodal values from $\text{Uniform}(0,1)$ is a modeling choice. Since $J_h$ is linear in the nodal values of the $q$-field, another distribution would multiply $\sigma_J$ by the same factor for all models and would therefore not change the ranking of the models by $\sigma_J$.}. The values on the inner and outer contours remain fixed at 1 and 0, respectively, so that every sampled field is continuous and piecewise linear with the prescribed contour values, and is therefore admissible (\sect{sec:jint}). \app{app:detailedbalance} proves that this update satisfies detailed balance, which, together with the fact that every interior node is eventually redrawn, ensures convergence to the stationary distribution. For efficiency, only the elemental contributions $J^e$ of the elements adjacent to the updated node are recomputed at each step, which avoids a full re-evaluation of the domain integral.

For each model, four independent chains are run, each initialized with a distinct seed of the MRG32k3a pseudorandom number generator~\cite{lecuyer:1999}, which makes every chain reproducible. In each chain, the first $N_{\text{burn}}$ steps are discarded as burn-in to remove the influence of the common starting field $q_{\text{lin}}$, and the following $N_{\text{sample}}$ steps are retained as samples. Once the agreement of the four chains has been verified (\sect{sec:convdiag}), their samples are combined, and the mean $\bar J$, the standard deviation $\sigma_J$, and the histograms reported below are computed from the combined set.

\begin{algorithm}[t!]
\caption{MCMC sampling of the $q$-field for uncertainty quantification of the $J$-integral (one chain)}
\begin{algorithmic}[1]
\State \textbf{Input:} Linearly interpolated initial $q$-field, $q_{\text{lin}}$
\State \textbf{Input:} Set of $N_{\text{int}}$ interior continuum nodes (excluding inner- and outer-contour nodes, where $q$ is fixed at 1 and 0, respectively)
\State \textbf{Input:} Precomputed elemental $J$-integral contributions $J^e$ computed with $q_{\text{lin}}$ for all $N_{\text{e}}$ elements in the integration domain
\State \textbf{Input:} Random seed, burn-in length $N_{\text{burn}}$, number of samples $N_{\text{sample}}$
\State \textbf{Output:} $J$-integral samples $\{J^{(k)}\}$, mean $\bar{J}$, standard deviation $\sigma_J$
\Statex
\State Initialize the random number generator with the seed
\State $q \gets q_{\text{lin}}$
\State $J \gets \sum_{e} J^e$ \Comment{Initial $J$-integral from linear $q$-field}
\For{$k = 1$ to $N_{\text{burn}} + N_{\text{sample}}$}
    \State Select node $i \sim \text{Uniform}\{1,\dots,N_{\text{int}}\}$ \Comment{Random interior node}
    \State Sample $(q^i)' \sim \text{Uniform}(0,1)$ \Comment{New admissible value}
    \State $q^i \gets (q^i)'$
    \For{each element $e$ adjacent to node $i$}
        \State $J \gets J - J^e$ \Comment{Remove stale elemental contribution}
        \State Recompute $(\nabla q)^e$ and $J^e$ with the updated $q^i$ using \eqn{eq:jele3}
        \State $J \gets J + J^e$ \Comment{Add updated elemental contribution}
    \EndFor
    \If{$k > N_{\text{burn}}$} \Comment{Discard burn-in}
        \State $J^{(k)} \gets J / B$ \Comment{Normalize by out-of-plane thickness}
        \State Update $\bar{J}$ and $\sigma_J$
    \EndIf
\EndFor
\State \Return $\{J^{(k)}\}$, $\bar{J}$, $\sigma_J$
\end{algorithmic}
\label{alg:mcmc_qfield}
\end{algorithm}

\subsection{Convergence diagnostics of the sampling}
\label{sec:convdiag}

Since only one of the $N_{\text{int}}$ nodal values changes per step, successive samples are strongly correlated. The decorrelation of the chain is characterized by the autocorrelation function $\rho(k)$, i.e.,\ the correlation between samples $k$ steps apart, together with the integrated autocorrelation time $\tau$ and the effective sample size $N_{\text{eff}}$~\cite{sokal:1997, gelman:carlin:2013},

\begin{equation}
\rho(k) = \frac{\operatorname{Cov}(J_m, J_{m+k})}{\operatorname{Var}(J)}, \qquad
\tau = 1 + 2\sum_{k=1}^{\infty}\rho(k), \qquad N_{\text{eff}} = \frac{N_{\text{sample}}}{\tau},
\label{eq:tau}
\end{equation}
where $J_m$ is the sampled value at step $m$. In the stationary state, $\rho(k)$ is independent of $m$, and $\tau$ is the number of steps required for the chain to correspond to an effectively independent sample.

\paragraph{Theoretical prediction.}
For the present sampler, $\rho(k)$ can be derived exactly. Since each $J^e$ in \eqn{eq:jele3} is linear in the nodal $q$-values, $J$ varies linearly with the interior nodal values. In the stationary state, a given nodal value remains unchanged after $k$ steps with probability $(1-1/N_{\text{int}})^k$, and is otherwise replaced by an independent draw that is uncorrelated with its previous value. The covariance between $J$ at steps $m$ and $m+k$ is therefore $(1-1/N_{\text{int}})^k$ times the variance of $J$ (see \app{app:acf}), which gives
\begin{equation}
\rho(k) = \left(1-\frac{1}{N_{\text{int}}}\right)^{k} \approx e^{-k/N_{\text{int}}}, \qquad \tau = 2N_{\text{int}} - 1.
\label{eq:rho}
\end{equation}
The autocorrelation thus depends only on $N_{\text{int}}$, and not on the continuum fields or the load, so that the diagnostics need to be evaluated only once per model; we evaluate them at $K_\rI = 0.1$~eV/\AA$^{5/2}$.

\paragraph{Choice of chain parameters.}
\eqn{eq:rho} allows the chain parameters to be set in advance. We use $N_{\text{burn}} \ge 28N_{\text{int}}$. After $N_{\text{burn}} = 28N_{\text{int}}$ steps (fourteen autocorrelation times), the expected fraction of nodes that still carry their initial value from $q_{\text{lin}}$ is $(1-1/N_{\text{int}})^{N_{\text{burn}}} < e^{-28}$. With $N_{\text{sample}} \ge 2000\,N_{\text{int}}$, each chain has $N_{\text{eff}} \ge 1000$, and the relative statistical error of the estimated $\sigma_J$, approximately $1/\sqrt{2N_{\text{eff}}}$, is at most about 2\%.

\paragraph{Estimation from the chains.}
To verify \eqn{eq:rho}, $\rho(k)$ is estimated for each chain from its post-burn-in sequence $J_1,\dots,J_{N_{\text{sample}}}$ as
\begin{equation}
\hat\rho(k) = \frac{\hat\gamma(k)}{\hat\gamma(0)}, \qquad
\hat\gamma(k) = \frac{1}{N_{\text{sample}}}\sum_{i=1}^{N_{\text{sample}}-k}\big(J_i-\bar J\big)\big(J_{i+k}-\bar J\big),
\label{eq:acf_est}
\end{equation}
where $\bar J$ is the chain mean, $\hat\gamma(k)$ is the estimated autocovariance at lag $k$, and $\hat\gamma(0)$ is the chain variance. The autocovariance $\hat\gamma(k)$ is evaluated for all lags at once by a zero-padded fast Fourier transform. Since $\hat\rho$ fluctuates at large lags due to statistical noise, the sum in \eqn{eq:tau} is truncated at the smallest lag $M$ satisfying $M \ge 5\hat\tau(M)$~\cite{sokal:1997}. The agreement between the four chains is measured by the Gelman--Rubin statistic $\hat R$~\cite{gelman:rubin:1992}, which compares the variance between the chain means with the variance within the chains and approaches 1 when all chains sample the same distribution (see \app{app:acf}).

\begin{figure}[t!]
    \centering
    \includegraphics[width=0.7\textwidth]{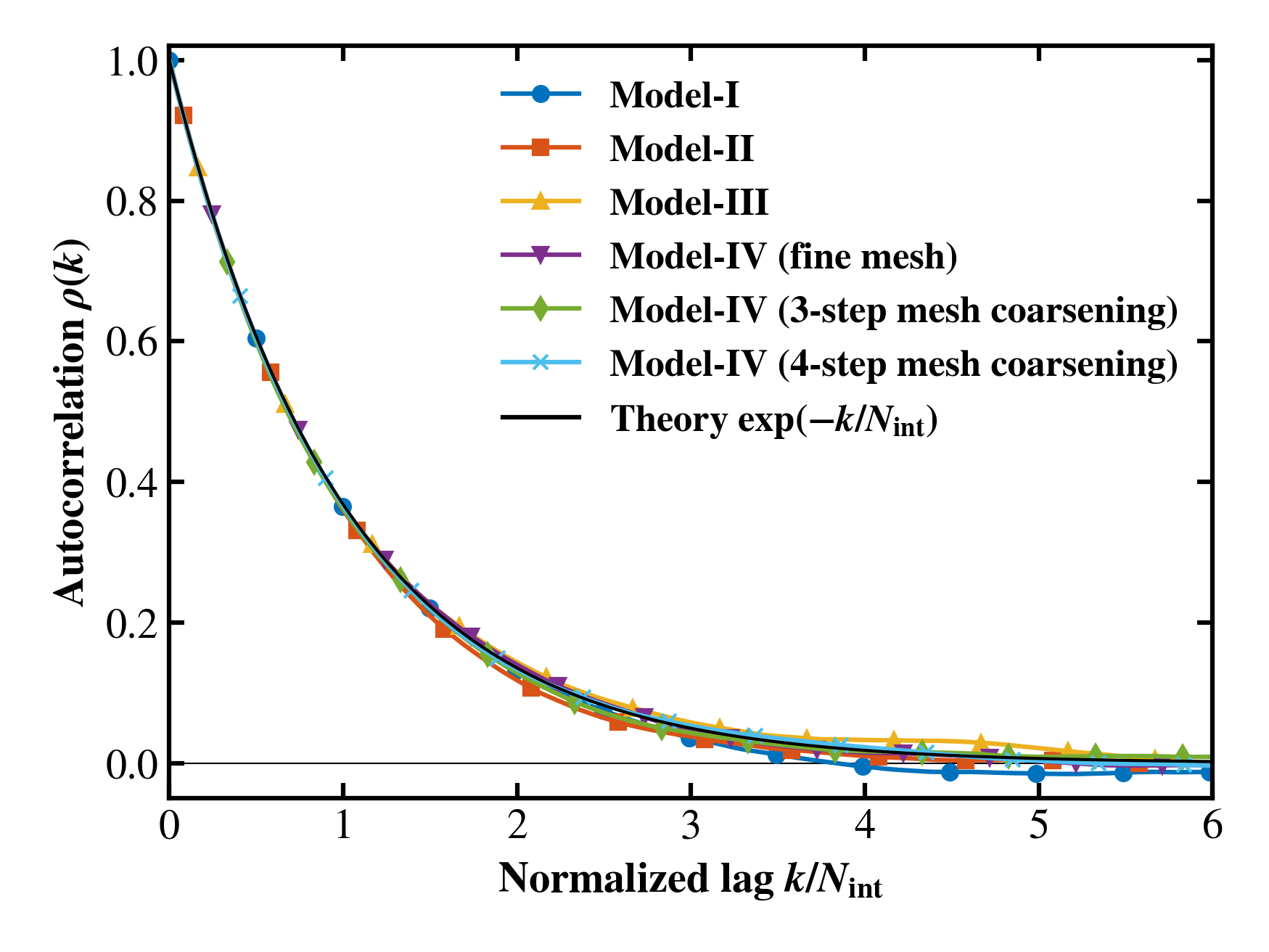}
    \caption{Autocorrelation of the sampled $J$-integral versus the normalized lag $k/N_{\text{int}}$ for all models in \tab{tab:mcmcdiag}, averaged over the four chains of each model. All models collapse onto the prediction $e^{-k/N_{\text{int}}}$ of \eqn{eq:rho} (black line).}
    \label{fig:acf}
\end{figure}

\paragraph{Results.}
\fig{fig:acf} compares the estimated autocorrelation, averaged over the four chains of each model, with \eqn{eq:rho}. When plotted against $k/N_{\text{int}}$, the curves for all six models, with $N_{\text{int}}$ ranging from 341 to 32{,}603, collapse onto the predicted curve. Since the prediction follows from the update rule alone, whereas $\hat\rho$ is computed from the sampled $J$ values alone, this agreement independently confirms that the chains decorrelate at the predicted rate. As listed in \tab{tab:mcmcdiag}, the predicted value $2N_{\text{int}}-1$ lies within the range of $\hat\tau$ obtained from the four chains for each model. In addition, each chain has $N_{\text{eff}} > 1000$, and $\hat R \le 1.0003$ for all models, well below the commonly used threshold of 1.01. The scatter of $\sigma_J$ between the four chains does not exceed 1.9\% for any model, consistent with the expected statistical error. The sampling results reported below are therefore converged.

\begin{table}[t]
\centering
\caption{Convergence diagnostics of the $q$-field sampling. $N_{\text{sample}}$: samples per chain after burn-in; $\hat\tau$: estimated integrated autocorrelation time (range over the four chains); min $N_{\text{eff}}$: smallest effective sample size of the four chains; $\hat R$: Gelman--Rubin statistic. For Model-IV, fine, 3-step, and 4-step denote the fine mesh and the meshes with three and four coarsening bands, respectively.}
\label{tab:mcmcdiag}
\begin{tabular}{ccccccc}
\hline
Model & $N_{\text{int}}$ & $N_{\text{sample}}$ & $2N_{\text{int}}-1$ ($\tau_{\text{pred}}$) & $\hat\tau$ & min $N_{\text{eff}}$ & $\hat R$ \\
\hline
Model-I          & \num{341}   & \num{1.00e6} & \num{681}   & \num{569}--\num{709}     & \num{1410} & \num{1.0003} \\
Model-II         & \num{1781}  & \num{4.45e6} & \num{3561}  & \num{2914}--\num{3641}   & \num{1222} & \num{1.0003} \\
Model-III        & \num{8115}  & \num{2.13e7} & \num{16229} & \num{14242}--\num{18788} & \num{1131} & \num{1.0002} \\
Model-IV, fine   & \num{32603} & \num{1.98e8} & \num{65205} & \num{63077}--\num{67977} & \num{2913} & \num{1.0002} \\
Model-IV, 3-step & \num{17095} & \num{6.93e7} & \num{34189} & \num{28701}--\num{39482} & \num{1755} & \num{1.0002} \\
Model-IV, 4-step & \num{7939}  & \num{6.93e7} & \num{15877} & \num{14673}--\num{16001} & \num{4331} & \num{1.0000} \\
\hline
\end{tabular}
\end{table}

\subsection{Effect of integration-domain placement}
\label{sec:domainplace}

\begin{figure}[t]\centering
     \begin{subfigure}[b]{1.0\textwidth}
         \centering
         \includegraphics[trim = 0mm 0mm 0mm 0mm, clip=true,width=1.0\textwidth]{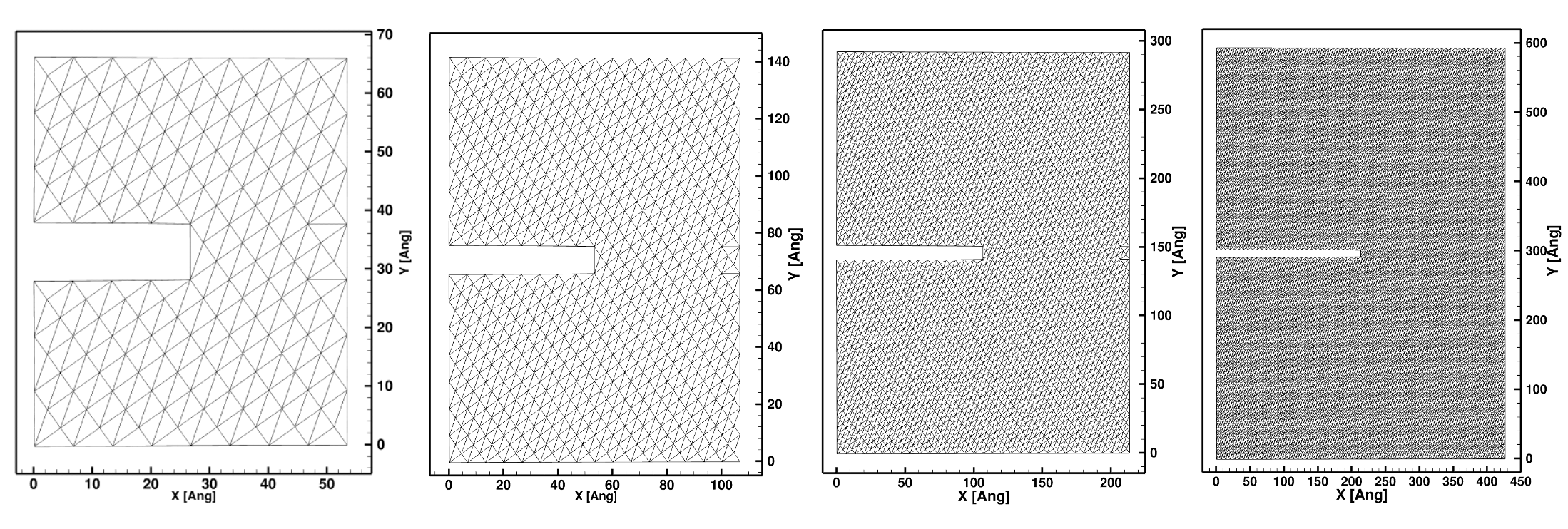}
         \caption{}
         \label{fig:fourmodels}
     \end{subfigure}
     \begin{subfigure}[b]{1.0\textwidth}
         \centering
         \includegraphics[trim = 0mm 0mm 0mm 0mm, clip=true,width=1.0\textwidth]{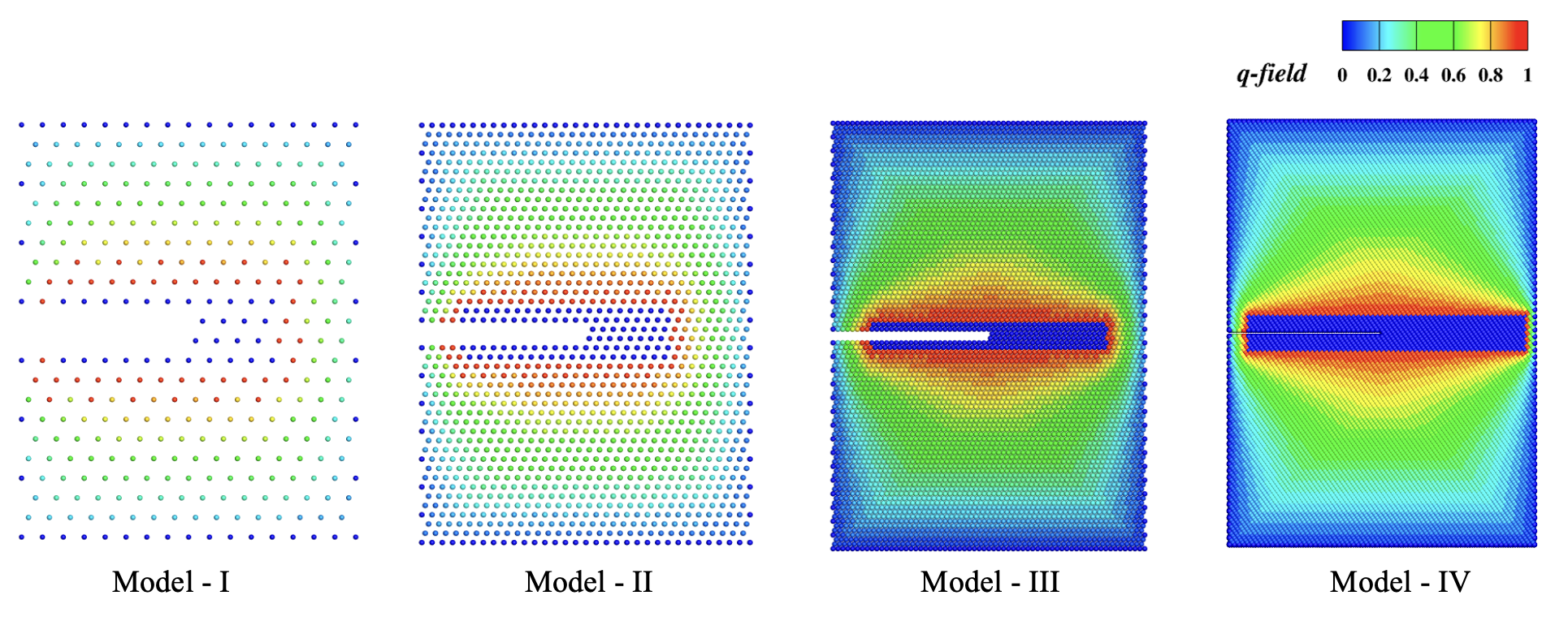}
         \caption{}
         \label{fig:fourqfields}
     \end{subfigure}
     \caption{(a) Reference mesh configurations of the four models; Models I-IV correspond to a successive ``zoom-out'' of the region around the crack tip. (b) Initial linearly interpolated $q$-fields of all models. The inner blue region ($q=0$) is atomistic and does not enter the $J$-integral calculation; in the continuum region, which forms the integration domain, $q$ varies from 1 to 0.}
     \label{fig:fourmodels_all}
\end{figure}

To examine how $\sigma_J$ responds to the placement of the integration domain, the sampling is applied to four models of increasing physical size but identical atomic-scale mesh resolution, shown in \fig{fig:fourmodels}. Both the size of the integration domain and its distance from the crack tip scale with the model, so that the domain moves progressively into the far-field; the corresponding initial $q$-fields are shown in \fig{fig:fourqfields}. Sampling is performed at $K_\rI = 0.1$~eV/\AA$^{5/2}$ (small-strain regime) and $K_\rI = 0.8$~eV/\AA$^{5/2}$ (finite-strain regime). Each sample is normalized by $K_\rI^2$, and the resulting distributions are shown as histograms with 200 bins.

\begin{figure}[t]
    \centering
    \includegraphics[width=\textwidth]{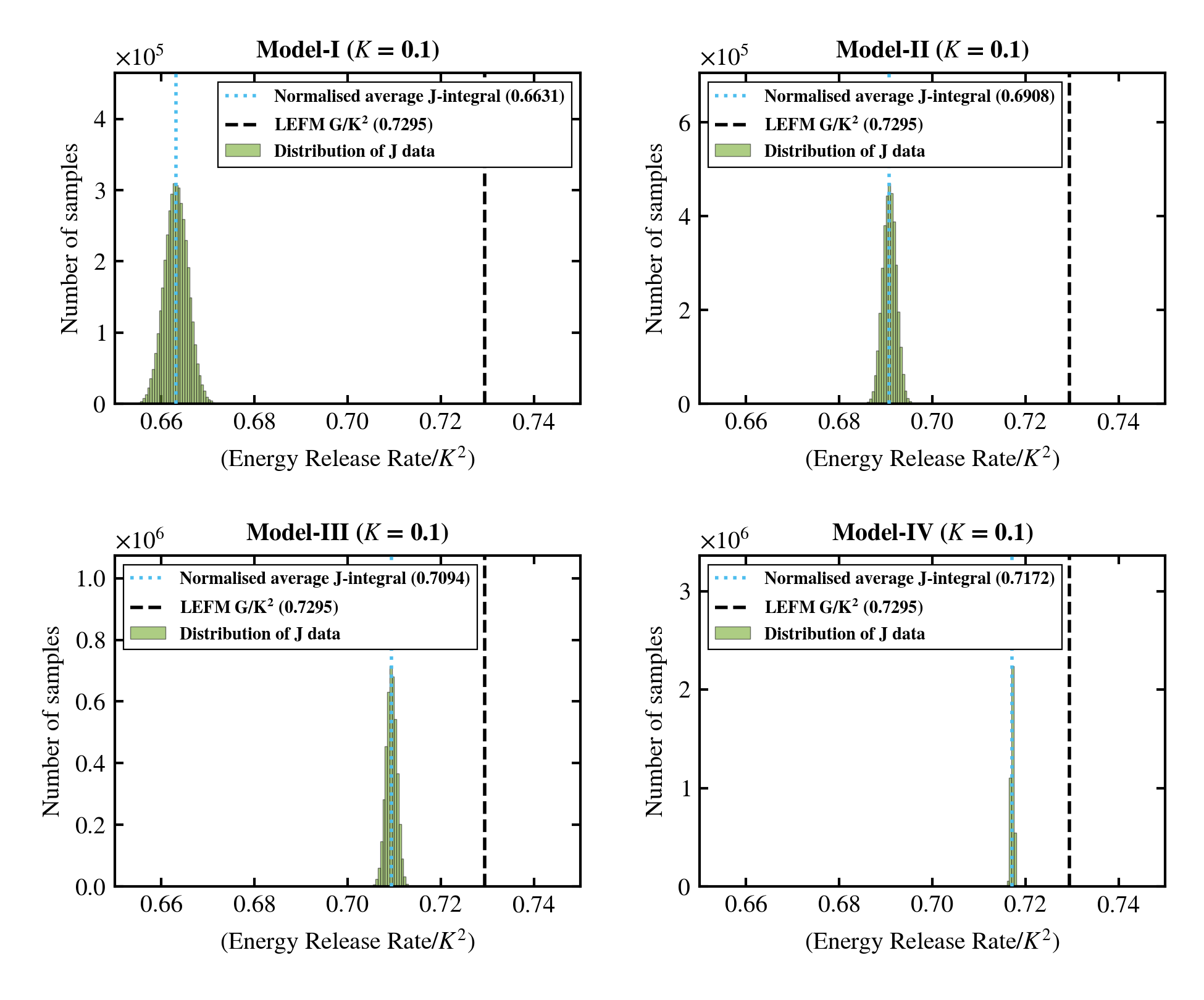}
    \caption{Distributions of the sampled $J$-integral values (normalized by $K_\rI^2$, four chains combined) for Models~I to IV at $K_\rI = 0.1$~eV/\AA$^{5/2}$.}
    \label{fig:ssbins}
\end{figure}

\begin{figure}[t]
    \centering
    \includegraphics[width=0.49\textwidth]{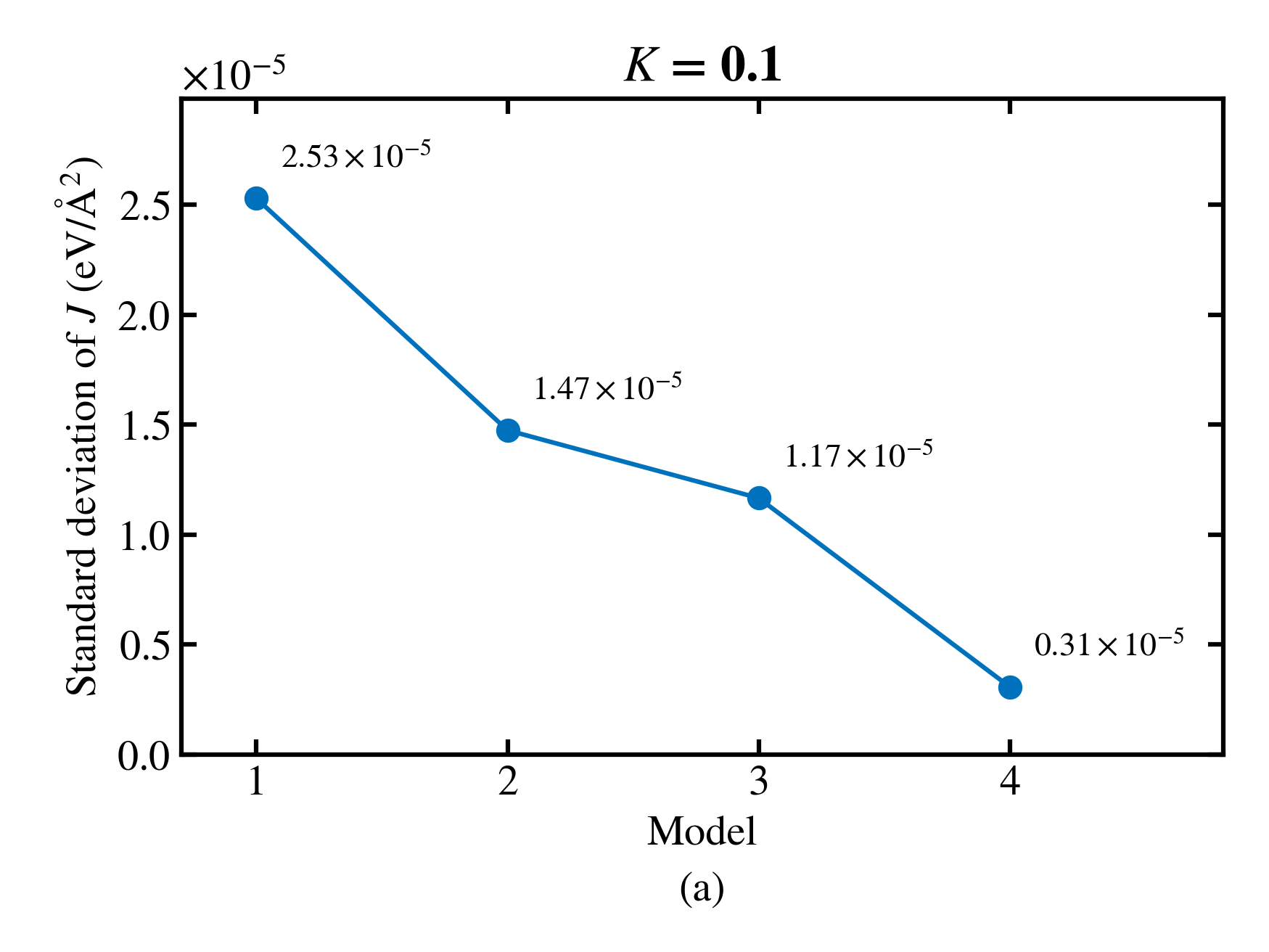}\hfill
    \includegraphics[width=0.49\textwidth]{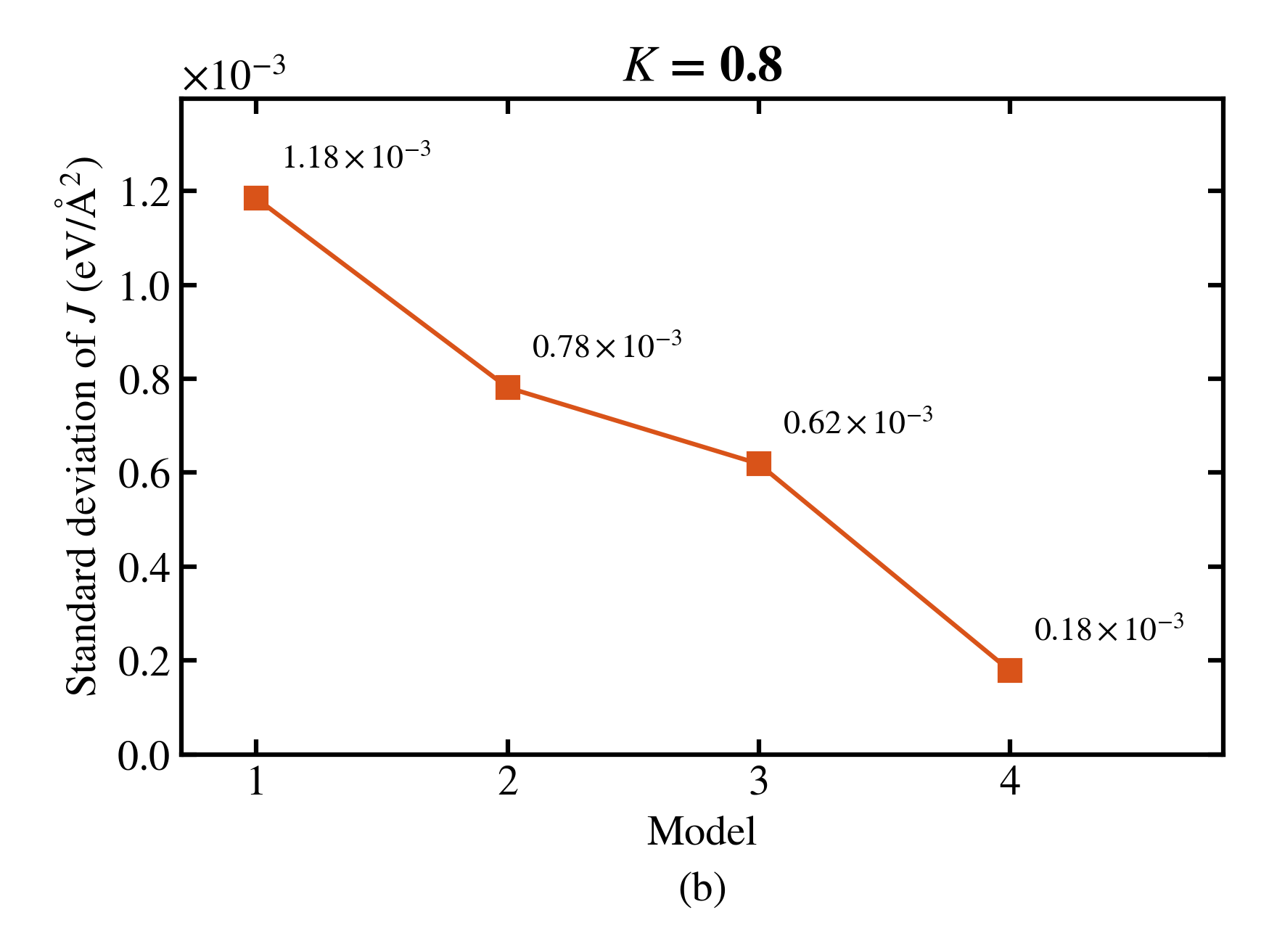}
    \caption{Standard deviation of the sampled $J$-integral (four chains combined) for Models~I to IV: (a) small-strain regime, $K_\rI = 0.1$~eV/\AA$^{5/2}$; (b) finite-strain regime, $K_\rI = 0.8$~eV/\AA$^{5/2}$.}
    \label{fig:std}
\end{figure}

\paragraph{Small-strain regime.}
In the small-strain regime, the applied $K$-field is expected to describe the fields away from the crack tip, and the corresponding LEFM energy release rate $G$ of a sharp crack in an infinite linear elastic body serves as the reference. In the smaller models, however, the fields in the integration domain are affected by the nonlinear crack-tip region and the nearby boundary; they therefore deviate from the $K$-field and vary steeply. As the model size increases, the near-tip disturbance decays before reaching the boundary, and the fields in the integration domain vary more gradually and approach the $K$-field. The discrete fields then satisfy configurational equilibrium more closely, which is reflected in a markedly narrower distribution of $J$ (\fig{fig:ssbins}) and a monotonic decrease in $\sigma_J$ by a factor of about eight, from $2.53\times10^{-5}$~eV/\AA$^2$ for Model-I to $0.31\times10^{-5}$~eV/\AA$^2$ for Model-IV (\fig{fig:std}(a)). At the same time, as $K$-dominance improves, the mean $\bar J/K_\rI^2$ approaches the LEFM value of 0.7295, with the deviation decreasing from $9.1\%$ below it for Model-I to $1.7\%$ for Model-IV. The decrease in $\sigma_J$ and the approach of $\bar J$ to $G$ thus both reflect the improvement of the continuum fields in the integration domain.

\begin{figure}[t]
    \centering
    \includegraphics[width=\textwidth]{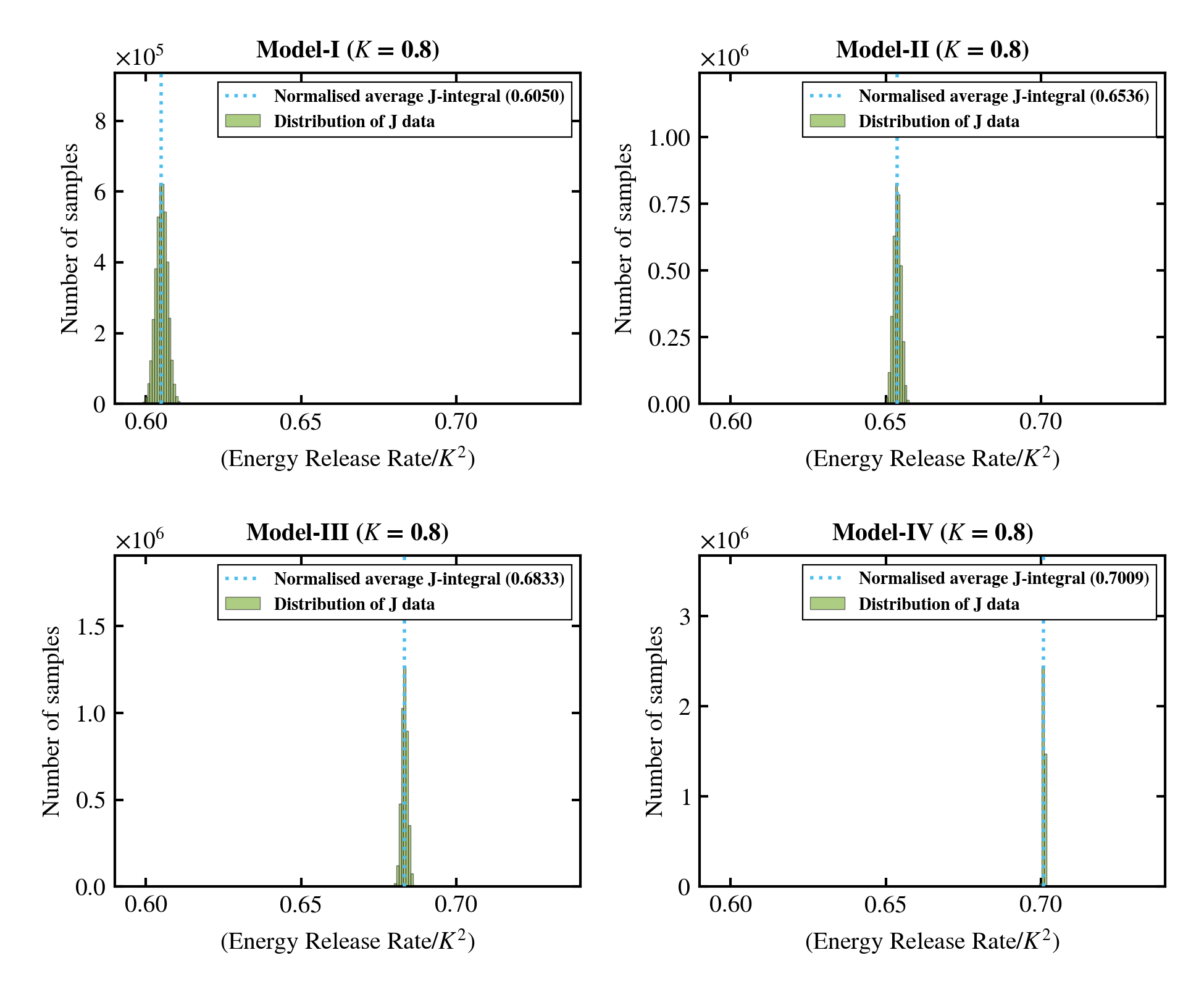}
    \caption{Distributions of the sampled $J$-integral values (normalized by $K_\rI^2$, four chains combined) for Models~I to IV at $K_\rI = 0.8$~eV/\AA$^{5/2}$.}
    \label{fig:fsbins}
\end{figure}

\paragraph{Finite-strain regime.}
The same analysis is carried out in the finite-strain regime at $K_\rI = 0.8$~eV/\AA$^{5/2}$. The behavior is the same as at small strain: as the model size increases and the integration domain moves away from the crack tip into regions where the fields vary more gradually, the distribution of the sampled $J$-integral values narrows (\fig{fig:fsbins}), and the standard deviation decreases monotonically by a factor of about seven, from $1.18\times10^{-3}$~eV/\AA$^2$ for Model-I to $0.18\times10^{-3}$~eV/\AA$^2$ for Model-IV (\fig{fig:std}(b)).

The results are consistent across the two loads: in both regimes, $\sigma_J$ decreases from Model-I to Model-IV in similar proportions (\fig{fig:std}). The $q$-field uncertainty therefore ranks the integration domains identically in the small-strain and finite-strain regimes.

\subsection{Effect of continuum mesh coarsening}
\label{sec:meshcoarsening}

\begin{figure}[t]
    \centering
    \includegraphics[width=\textwidth]{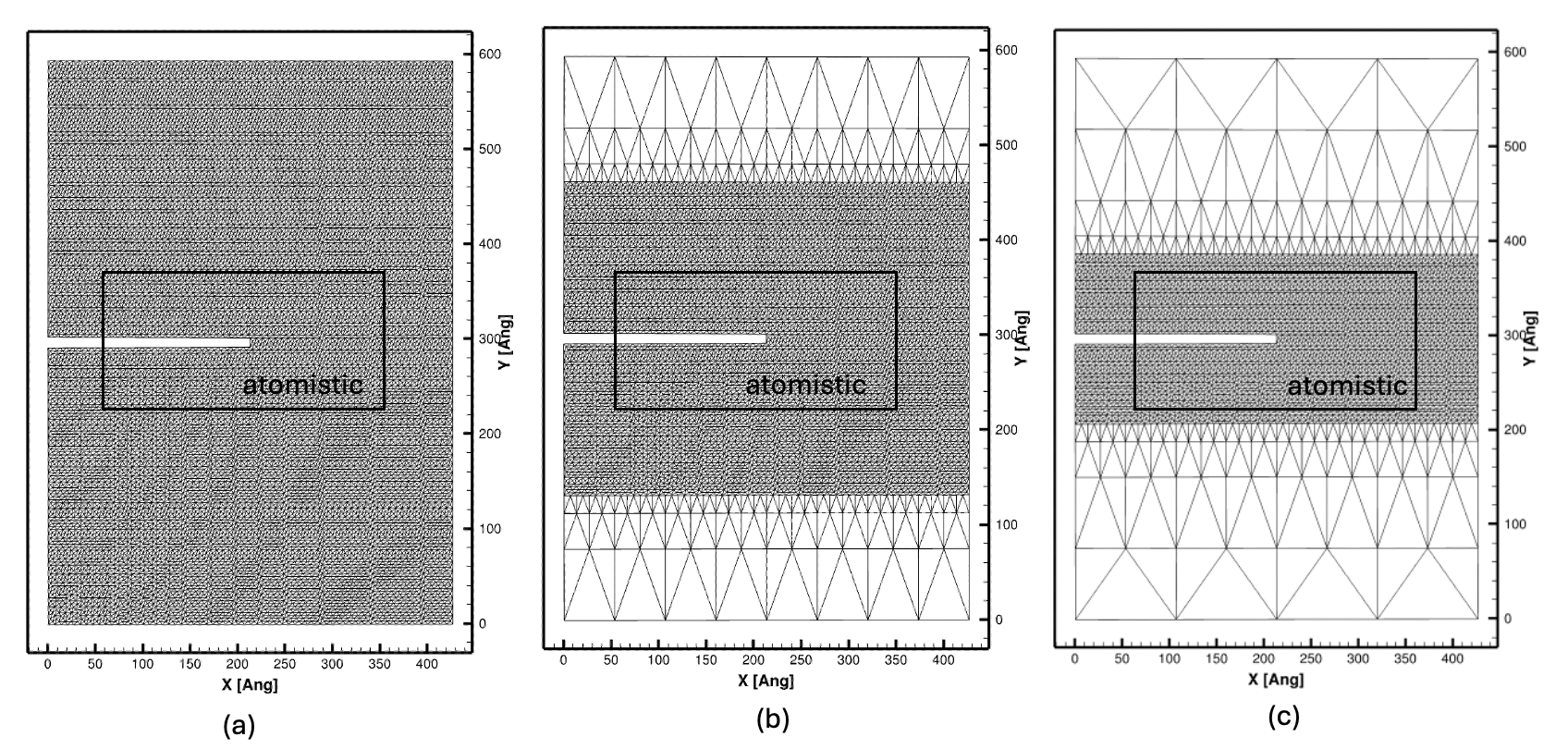}
    \caption{Continuum meshes of Model-IV used in the mesh-coarsening study: (a) fine mesh, (b) 3-step coarsening, and (c) 4-step coarsening, with three and four coarsening bands on each side of the fully refined strip. The atomistic region, the integration domain and the outer boundary are identical in all three meshes.}
    \label{fig:coarsemeshes}
\end{figure}

In practical QC simulations, the continuum region is coarse-grained away from the crack tip. To examine the effect of coarse-graining, the continuum mesh of Model-IV is coarsened in three or four bands above and below a fully refined strip around the crack plane, with the element size doubling from one band to the next (\fig{fig:coarsemeshes}). The atomistic region, the integration domain, and the loading remain unchanged, while coarsening reduces the number of repatoms from 36{,}416 (fine) to 20{,}468 (3-step) and 11{,}240 (4-step). Since the three meshes model the same problem, the fine-mesh result serves as the reference, and any change in $\bar J$ is due to the coarsening error alone.

\begin{figure}[t]
    \centering
    \includegraphics[width=\textwidth]{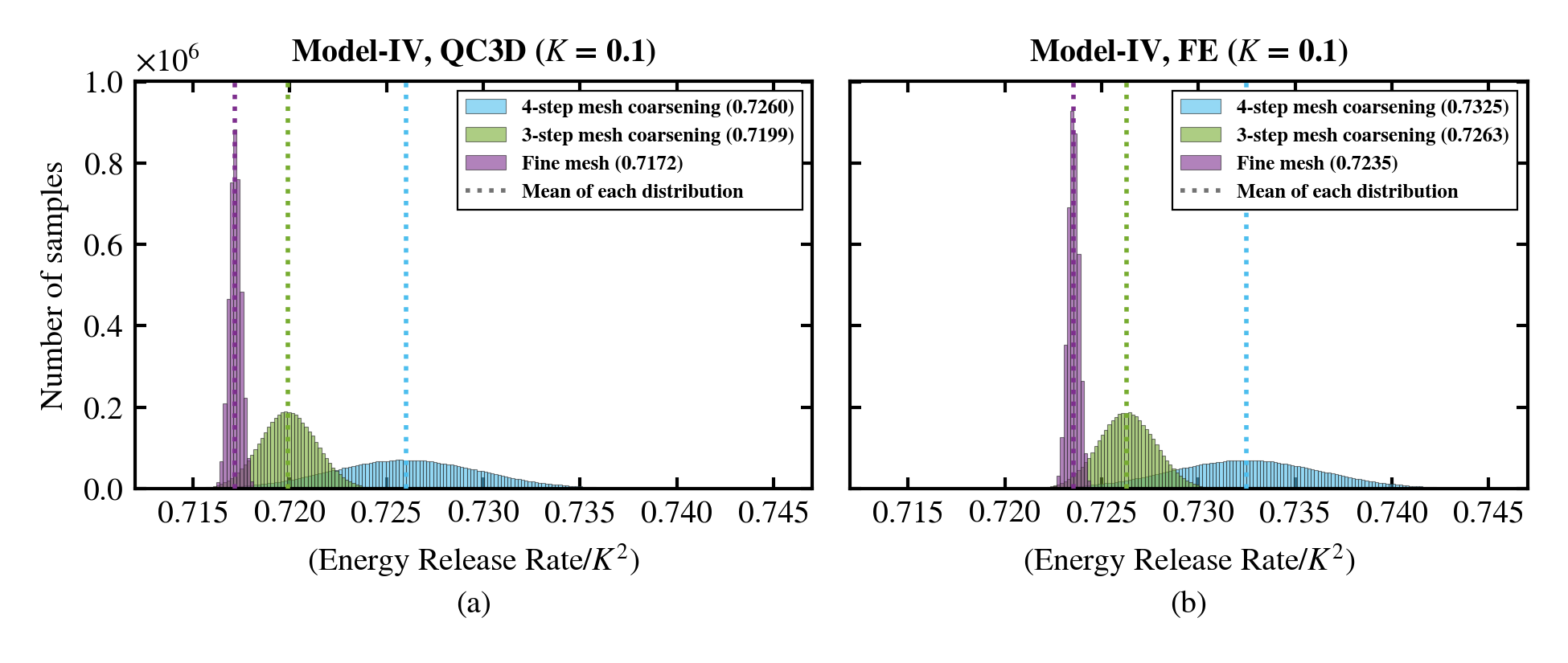}
    \caption{Distributions of the sampled $J$-integral values (normalized by $K_\rI^2$, four chains combined) for Model-IV with the fine mesh and with 3-step and 4-step mesh coarsening at $K_\rI = 0.1$~eV/\AA$^{5/2}$: (a) QC3D, (b) conventional FE (fully local model). Dotted lines mark the mean of each distribution.}
    \label{fig:coarsebins}
\end{figure}

The results at $K_\rI = 0.1$~eV/\AA$^{5/2}$ are shown in \fig{fig:coarsebins}(a) and \tab{tab:coarsening}. The mean is only weakly affected: although the number of repatoms is reduced by a factor of about three, $\bar J$ departs from the fine-mesh value by only $0.38\%$ (3-step) and $1.23\%$ (4-step), which supports the use of coarse-grained multiscale simulations. The $q$-field uncertainty increases with coarsening: relative to the fine mesh, $\sigma_J$ increases by factors of 4.7 and 12.6 for 3-step and 4-step coarsening, respectively. As seen in \fig{fig:coarsebins}(a), the coarsening error of each coarsened mesh is approximately twice its $\sigma_J$. Coarsening thus degrades the fields in the integration domain; $\sigma_J$ reflects the resulting discretization error and correctly ranks the meshes by quality.

\begin{table}[t]
\centering
\caption{Effect of continuum mesh coarsening (Model-IV, $K_\rI = 0.1$~eV/\AA$^{5/2}$): number of repatoms, and mean and standard deviation of the sampled $J$-integral (normalized by $K_\rI^2$) for QC3D and for the conventional FE model.}
\label{tab:coarsening}
\begin{tabular}{cccccc}
\hline
 & & \multicolumn{2}{c}{QC3D} & \multicolumn{2}{c}{FE} \\
\cline{3-4}\cline{5-6}
Mesh & Repatoms & $\bar J/K_\rI^2$ & $\sigma_J/K_\rI^2$ & $\bar J/K_\rI^2$ & $\sigma_J/K_\rI^2$ \\
\hline
Fine              & \num{36416} & \num{0.7172} & \num{0.31e-3} & \num{0.7235} & \num{0.29e-3} \\
3-step coarsening & \num{20468} & \num{0.7199} & \num{1.46e-3} & \num{0.7263} & \num{1.47e-3} \\
4-step coarsening & \num{11240} & \num{0.7260} & \num{3.91e-3} & \num{0.7325} & \num{3.94e-3} \\
\hline
\end{tabular}
\end{table}

In both studies, $\sigma_J$ decreases as the integration domain moves away from the crack tip and increases as the mesh is coarsened and thus provides a reliable indicator of the quality of the integration domain and mesh without requiring a reference solution. Since the integration domain in QC is an FE-discretized continuum region, and $J$ and $\sigma_J$ are computed solely from its FE fields, the UQ procedure should apply equally to conventional FE calculations, as demonstrated next.

\subsection{Demonstration on a conventional FE calculation}
\label{sec:fedemo}

To demonstrate the sampling framework on a conventional FE calculation, the mesh-coarsening study is repeated with the nonlocal (atomistic) description of QC3D switched off. The region inside the inner contour, previously atomistic, is then modeled by the same Cauchy--Born continuum as the rest of the model, so that the whole model reduces to a standard nonlinear FE model with linear tetrahedral elements and a hyperelastic constitutive law given by the Cauchy--Born rule. This is the only difference from the QC3D calculations: the three meshes, the loading, the relaxation, the integration domain, and the number of interior nodes are identical to those of \sect{sec:meshcoarsening}. The sampling uses the same chain parameters, and all chains have converged ($\hat R \le 1.0002$ and $N_{\text{eff}} > 1700$ per chain).

\fig{fig:coarsebins}(b) and \tab{tab:coarsening} show that the FE results reproduce the behavior observed with QC3D. For each mesh, $\sigma_J$ agrees with the QC3D value to within $7\%$ and increases with coarsening in the same way. The coarsening error is $0.38\%$ and $1.24\%$ for 3-step and 4-step coarsening, respectively, practically identical to the QC3D values, and again amounts to about $2\sigma_J$. The only difference in the results between the two sets of results is a nearly constant offset of the mean: $\bar J/K_\rI^2$ is higher in FE by about $0.0064$ for all three meshes. This offset reflects the different descriptions of the crack tip, which is atomistic in QC3D and a Cauchy--Born continuum in FE; it shifts $J$ by the same amount for every $q$-field and therefore does not contribute to $\sigma_J$.\footnote{The main difference between the approaches is that with increasing load, fracture through bond breaking would occur at the crack tip in the QC3D simulation, but not in the continuum Cauchy--Born model.}

The $q$-field uncertainty is therefore governed by the discretization of the integration domain alone and is insensitive to how the crack-tip region is modeled. The framework thus applies directly to conventional FE fracture calculations, where it quantifies the precision of $J$ and ranks meshes and integration domains by quality.

\section{Application to a three-point bending test}
\label{sec:application}

The validation under controlled $K$-field loading established the numerical correctness of the multiscale $J$-integral implementation in idealized conditions. As a demonstration under structural loading without a prescribed $K$-field, we apply the implementation to a three-point bending test, in which the energy release rate is not known a priori and the stress field results from the specimen geometry, supports, and applied load rather than being prescribed.

\begin{figure}[t!]
    \centering
    \includegraphics[trim = 0mm 0mm 0mm 0mm, clip=true, width=0.7
    \textwidth]{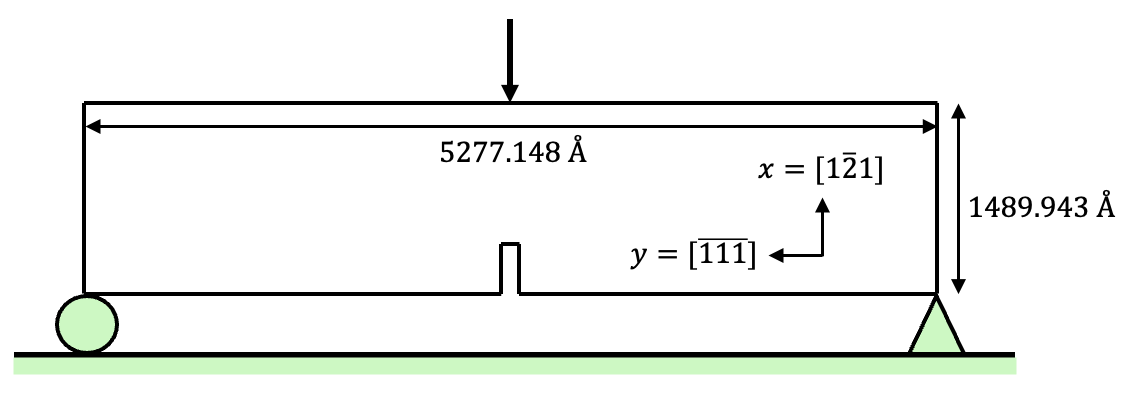}
    \caption{Schematic of the three-point bending specimen.}
    \label{fig:threeptbend}
\end{figure}

\begin{figure}[t!]
    \centering
    \begin{subfigure}[b]{0.7\textwidth}
        \centering
        \includegraphics[trim = 0mm 0mm 0mm 0mm, clip=true, width=\textwidth]{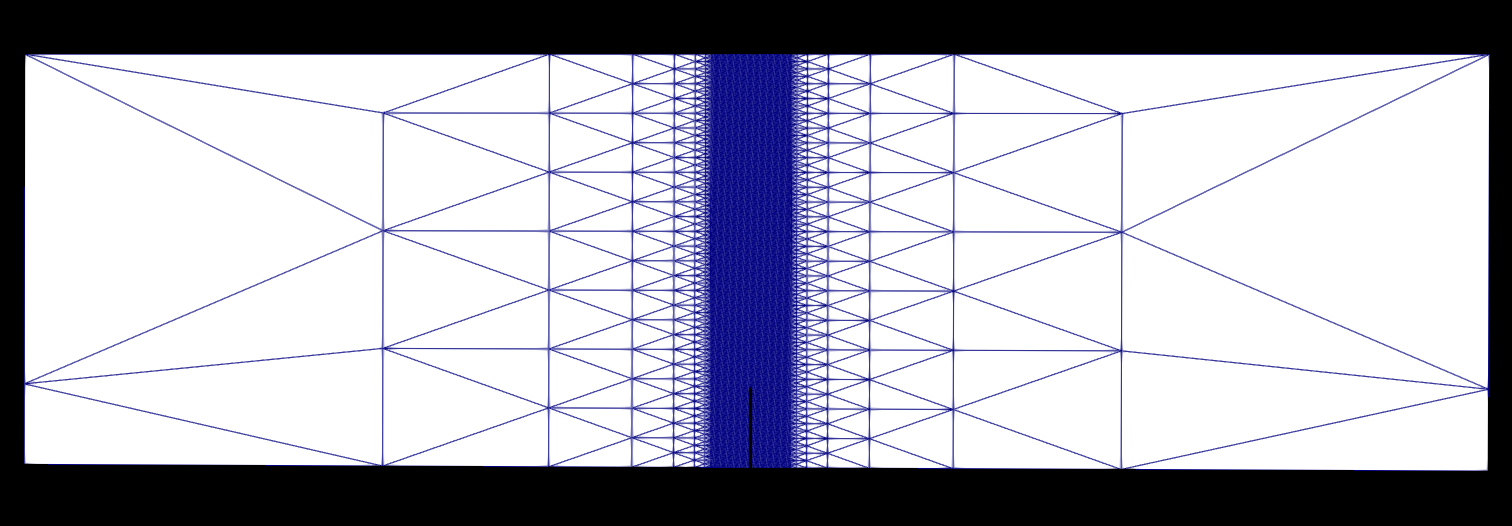}
        \caption{QC3D model}
        \label{fig:qc3d_atoms}
    \end{subfigure}
    \\[1em]
    \begin{subfigure}[b]{0.7\textwidth}
        \centering
        \includegraphics[trim = 0mm 0mm 0mm 0mm, clip=true, width=\textwidth]{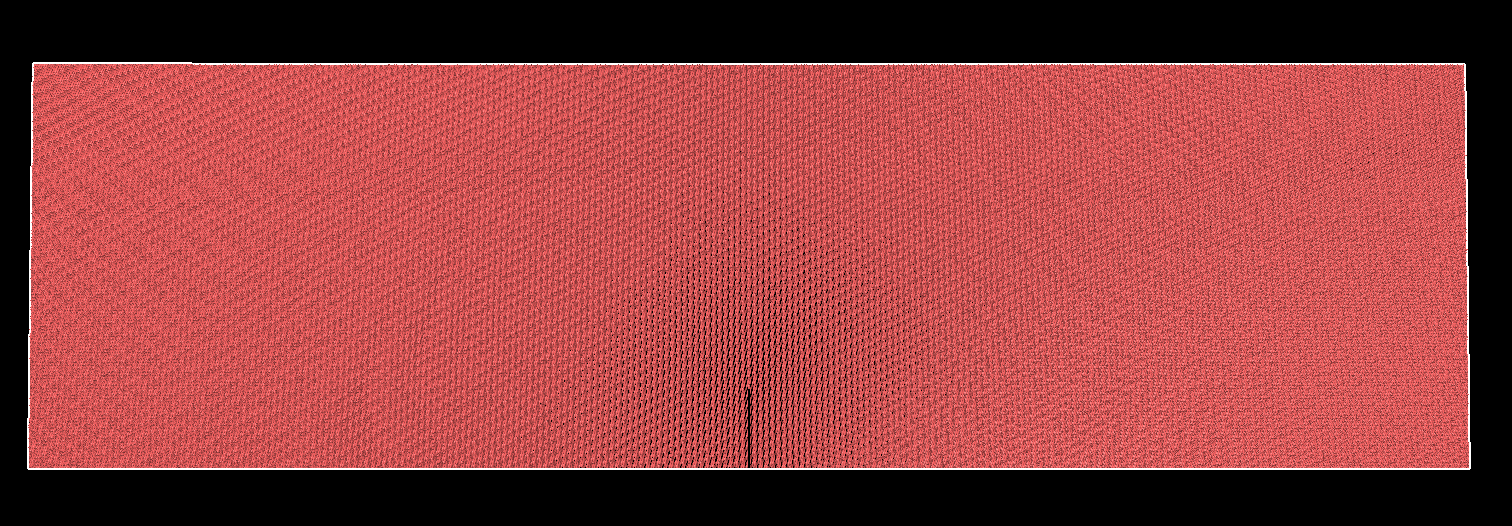}
        \caption{Fully atomistic model}
        \label{fig:fullatom_atoms}
    \end{subfigure}
    \caption{Comparison of the QC3D and fully atomistic models of the three-point bending specimen.}
    \label{fig:qc_vs_atomistic}
\end{figure}

We perform a QC3D simulation of a precracked single-crystal silicon three-point bending specimen, shown in \fig{fig:threeptbend}. Atomic interactions are described by the modified Stillinger--Weber three-body potential described in \sect{sec:cracl}. The model is periodic in the $z$ direction with a fixed width, which imposes plane-strain conditions. The silicon is oriented such that $x$ lies along $[1\bar21]$ and $y$ along $[111]$ crystallographic directions so that the crack propagates on the (111) cleavage plane. The dimensions of the specimen are $\SI{1489.943}{\angstrom}\times\SI{5277.148}{\angstrom}\times\SI{30.722}{\angstrom}$. Simulating a model of this size is made feasible by QC3D's coarse-graining. The QC3D model employs $358{,}452$ repatoms, in contrast to the $12{,}053{,}441$ atoms required for an equivalent fully atomistic simulation of the same specimen dimensions and crystal orientation (see \fig{fig:qc_vs_atomistic})---a $34$-fold reduction in the number of degrees of freedom, while retaining full atomistic resolution in the vicinity of the crack tip, where it is most needed. The edge crack has an opening of \SI{9.407}{\angstrom} and extends \SI{292.67}{\angstrom} into the specimen.

\begin{figure}[t!]
    \centering
    \includegraphics[trim = 0mm 0mm 0mm 0mm, clip=true, width=0.8
    \textwidth]{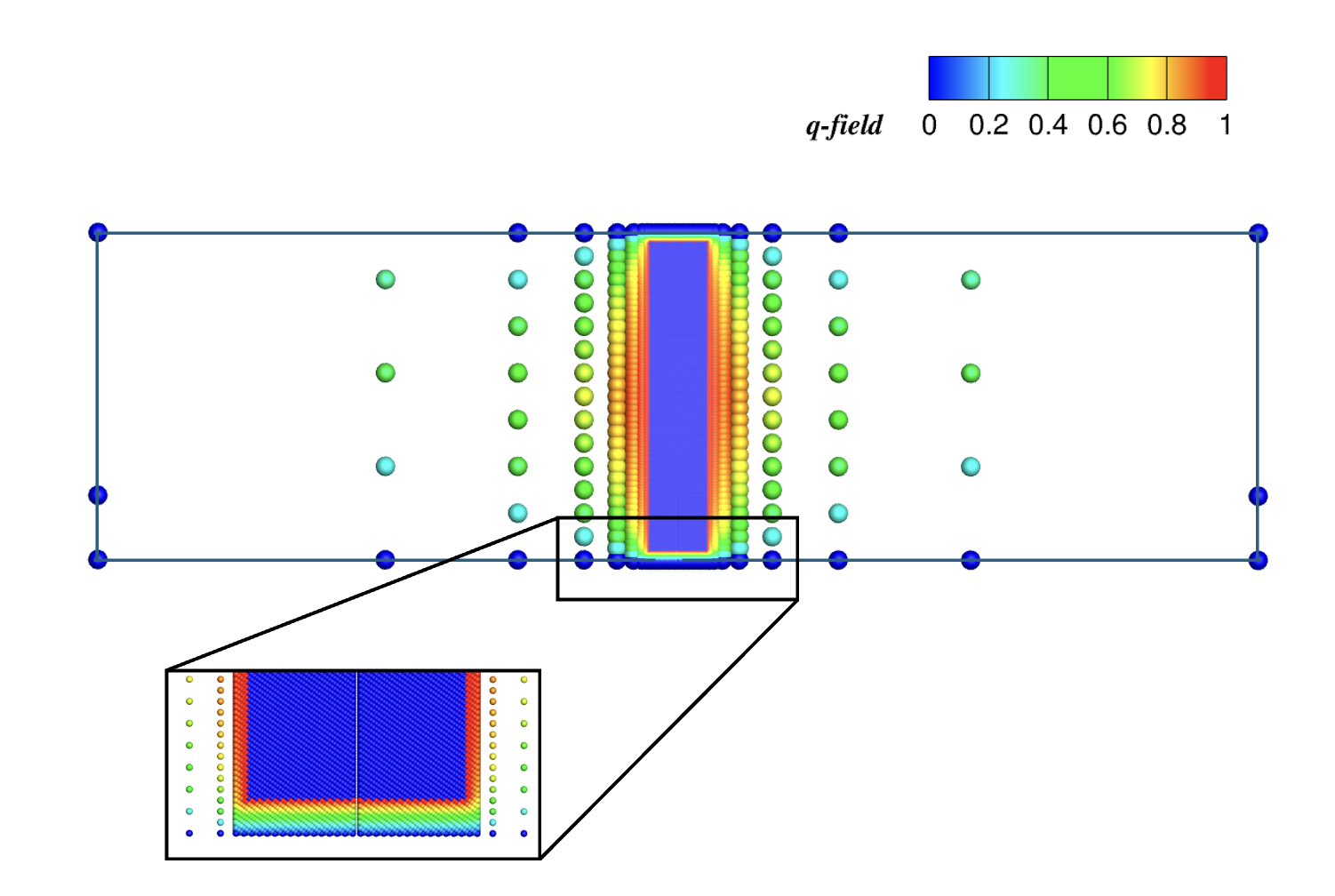}
    \caption{Linearly interpolated $q$-field of the three-point bending configuration. As in \fig{fig:fourqfields}, the internal blue region ($q=0$) marks the atomistic region excluded from the $J$-integral; $q$ decreases linearly from 1 to 0 across the continuum integration domain.}
    \label{fig:3ptbendqfield}
\end{figure}

\begin{table}[t]
\centering
\caption{Critical energy release rate at cleavage for the three-point bending test of single-crystal silicon modeled with the modified Stillinger--Weber potential: Griffith value $2\gamma_{\rm s}$ versus computed $J_{\rm c}$.}
\label{tab:j_validation}
\begin{tabular}{ccc}
\hline
$G_{\rm c} = 2\gamma_{\rm s}$ (\si{\electronvolt\per\square\angstrom}) & $J_{\rm c}$ (\si{\electronvolt\per\square\angstrom}) & Relative difference \\
\hline
\num{0.1814} & \num{0.1886} & \SI{3.85}{\percent} \\
\hline
\end{tabular}
\end{table}

The specimen is supported at its two ends, $y=0$ and $y=\SI{5277.148}{\angstrom}$, on the face $x=0$: the corner node at $y=0$ is fixed in all directions, and the one at the opposite end is fixed in $x$ and $z$ but free in $y$. The load is applied under force control on the opposite face ($x=\SI{1489.943}{\angstrom}$), as a force in the $-x$ direction distributed equally among the atoms in a strip one repeat cell (\SI{9.407} {\angstrom}) wide at mid-span, directly opposite the crack. The total force is increased by \SI{1.80}{\electronvolt\per\angstrom} per load step. 

At each load step, the model is statically relaxed through Jacobi-preconditioned conjugate-gradient energy minimization. The load is incrementally increased until crack propagation is detected by the cleavage-detection algorithm (\alg{alg:cleavage_convergence}). The $J$-integral is computed before the onset of unstable cleavage, using all continuum (local) elements as the integration domain and the linearly interpolated $q$-field shown in \fig{fig:3ptbendqfield}.

The results are summarized in \tab{tab:j_validation}. The $J$-integral at the last stable load step is compared with the Griffith value for ideally brittle cleavage. For single-crystal silicon under static conditions at zero temperature and under small-scale yielding conditions, the critical energy release rate should satisfy the Griffith energy balance, $J_{c} = 2\gamma_{\rm s}$, where $\gamma_{\rm s}$ is the surface energy of the (111) cleavage plane. The surface energy $\gamma_{\rm s}$ is computed with the same interatomic potential from a crystal that is periodic in all directions and about 1880~\AA{} thick in the [111] direction, as $\gamma_{\rm s} = (E_{\rm slab} - E_{\rm bulk})/(2A)$, where $E_{\rm bulk}$ is the relaxed energy of the periodic crystal and $E_{\rm slab}$ is the relaxed energy after the periodicity in the [111] direction is removed. Removing the periodicity creates two free (111) surfaces in vacuum, of area $A$, at the shuffle plane, the preferred cleavage plane of silicon, which yields $\gamma_{\rm s} = 0.0907$~eV/\AA$^2$. 

Our results show excellent agreement between the computed $J$-integral and $2\gamma_{\rm s}$, with a deviation of less than 4\%. Since the load is increased in steps of about 1.2\% of the critical load, and $J$ scales with the square of the load, $J$ increases by about 2.4\% over one step; so the true critical value exceeds $2\gamma_{\rm s}$ by 4–6.5\%. This excess is consistent with lattice trapping, which raises the load at which a crack becomes unstable above the Griffith value. This agreement shows that the $J$-integral implementation in QC3D extracts a consistent fracture driving force under structural loading without a prescribed $K$-field.

\section{Conclusions}
\label{sec:conclusions}

This work presents the first rigorous implementation and validation of the $J$-integral within a 3D quasicontinuum framework. Unlike purely atomistic $J$-integral formulations, it uses the continuum fields that are naturally available in QC: the stresses and strain energy densities obtained from the Cauchy--Born rule in the continuum region. In contrast to conventional FE implementations of the domain $J$-integral, which rely on phenomenological constitutive models, our formulation inherits its constitutive response directly from the underlying interatomic potential, ensuring that the computed energy release rate is consistent with the atomistic-scale physics governing fracture at the crack tip. In addition, this work adapts the VCE method to a concurrent atomistic--continuum multiscale framework in which relaxation of the atomistic degrees of freedom precludes the conventional single-mesh nodal perturbation procedure.

The $J$-integral implementation is comprehensively verified and validated against both LEFM theory and the VCE method in three regimes of increasing physical fidelity: first under idealized linear elastic conditions, then with the full nonlinear Cauchy--Born response, and finally with atomic relaxation enabled to allow the crack-tip region to reach a nonlinear equilibrium state. In all cases, the $J$-integral and VCE results agree closely, with a maximum deviation of about $3\%$ for the relaxed QC configuration (attributed to the numerical differentiation in the VCE approach). This close agreement is particularly notable given that the two methods are constructed independently---the $J$-integral from a domain integral of continuum fields, and the VCE method from a finite-difference energy comparison between two independently relaxed atomistic-continuum configurations---thus providing mutual validation of both approaches at finite deformation.

For exact continuum fields, the domain-form $J$-integral is independent of the choice of the weighting field $q$. For discrete fields, it varies with $q$ through a residual that arises solely from the discretization of the fields in the integration domain. Treating $q$ as the uncertain input and $J$ as the quantity of interest, we introduce an MCMC framework that samples the admissible $q$-fields and quantifies the resulting spread of $J$ by its standard deviation $\sigma_J$. This $q$-field uncertainty measures the precision of the computed $J$ with respect to the choice of $q$. The convergence of the sampling is verified against an exact prediction of its autocorrelation. Moving the integration domain away from the crack tip reduces $\sigma_J$ by a factor of about seven to eight, whereas coarsening the continuum mesh increases it by a factor of up to about 13, with the mean $J$ changing by at most 1.2\%. A conventional FE model gives practically the same $\sigma_J$ and coarsening error as QC3D and differs only by a shift of the mean $J$, confirming that the $q$-field uncertainty originates from the discretization of the integration domain and not from the modeling of the region inside the inner contour. In all cases studied, a lower $\sigma_J$ coincides with a more accurate $J$. The $q$-field uncertainty can therefore serve as a practical proxy for the accuracy of $J$ with respect to the discretization of the integration domain. It does not, however, capture effects that shift $J$ equally for every $q$-field, such as those arising from the modeling of the problem rather than its discretization (e.g., the nonlinear crack-tip response, the finite crack opening, or the finite model size). $\sigma_J$ thus provides a reliable indicator of the quality of the integration domain and mesh, obtained from a single simulation without requiring a refined reference mesh or analytical solution, for both multiscale and conventional FE fracture calculations.

The predictive capability of the approach for cases in which the applied stress intensity factor is unknown is demonstrated through a QC3D simulation of a three-point bending test of single-crystal silicon. The computed critical energy release rate is shown to be in excellent agreement with the Griffith criterion ($<4\%$ deviation), despite the complex, nonuniform stress field generated by this structural loading configuration. This confirms the method's ability to extract accurate fracture driving forces beyond idealized $K$-field boundary conditions and demonstrates that QC3D's coarse-graining---requiring only about 350k repatoms, compared with over $12$ million atoms for an equivalent fully atomistic model---makes such structural-scale fracture simulations computationally tractable while retaining full atomistic resolution at the crack tip.

The significance of this development is that it enables large-scale fracture simulations that directly link atomistic mechanisms around the crack tip to continuum-scale driving forces, bridging a critical gap in computational fracture mechanics. Unlike the VCE method, which requires two independently relaxed simulations per load step and is therefore comparatively costly, the $J$-integral can be evaluated directly from a single equilibrated configuration, making it the more practical choice.

Several avenues remain for future investigation. The interatomic potential used here is selected for computational efficiency rather than physical realism, and future work should apply this framework with more accurate potentials to study fracture in materials of genuine engineering interest, including shape-memory alloys, high-entropy alloys, and other complex multicomponent systems under realistic loading conditions. The current implementation targets straight cracks under quasistatic, athermal, plane-strain conditions; extending the formulation to curved or kinked crack fronts, dynamic crack propagation, and finite-temperature simulations via Hot-QC would broaden its applicability to a wider range of fracture problems. Finally, coupling the present framework with QC's automatic mesh adaption capability would allow the $J$-integral to be evaluated continuously as the atomistic region evolves with an advancing crack, rather than at fixed load steps with a static mesh, further enhancing the method's utility for large-scale fracture simulations.

\section*{Acknowledgements}
The authors acknowledge partial support by the National Science Foundation (NSF) under Awards CMMI-2428667 and DMR-1834251. The authors thank Woo Kyun Kim for valuable discussions and insights.

\appendix
\section{Conversion of the contour form of the $J$-integral into the domain form}
\label{app:contour-to-domain}

This appendix derives the domain form of the $J$-integral \cite{li:shih:1985} from Rice's contour form, retaining the term that vanishes for exact fields, in order to show how the $q$-dependence of the computed $J$ arises in numerical calculations. The derivation is presented in two dimensions; the plane-strain volume form, \eqn{eq:jdomain}, in the same way, with an additional integration through the thickness $B$ and division by $B$.

\subsection{Derivation}
\label{app:derivation}

The contour form of the $J$-integral, \eqn{eq:jcontour}, can be written in terms of the Eshelby tensor of \eqn{eq:eshelby} as
\begin{equation}
J^\Gamma = \int_{\Gamma} H_{1J}\, N_J \, d\Gamma,
\qquad
H_{1J} = W \delta_{1J} - P_{kJ} \frac{\partial u_k}{\partial X_1},
\label{eq:jcontour2}
\end{equation}
where $\Gamma$ is an open counterclockwise contour that runs from the lower to the upper crack face (\fig{contourint}), and $\bN$ is its unit normal pointing away from the crack tip.

Since the divergence theorem requires a closed contour, we complete $\Gamma$ with the crack-face segments $C_+$ and $C_-$ and an outer contour $C_0$, forming the closed contour $C = C_0 + C_+ - \Gamma + C_-$ that encloses the area $A_0$ (\fig{areaint}). Let $\boldsymbol{M}$ denote the unit outward normal to $A_0$ on $C$, and let $q(\bX)$ be a continuous, piecewise-differentiable scalar field that equals $1$ on $\Gamma$ and $0$ on $C_0$. The closed-contour integral of $q H_{1J}$ splits into four parts:
\begin{equation}
\oint_{C} q\, H_{1J}\, M_J \, dC
= \int_{C_0} q\, H_{1J}\, M_J \, dC
+ \int_{C_+} q\, H_{1J}\, M_J \, dC
+ \int_{C_-} q\, H_{1J}\, M_J \, dC
+ \int_{\Gamma} q\, H_{1J}\, M_J \, dC .
\label{eq:jcontour4}
\end{equation}
The integral over $C_0$ vanishes because $q = 0$ there. The crack faces are flat and traction-free ($M_1 = 0$ and $P_{kJ} M_J = 0$), so $H_{1J} M_J = 0$ on $C_\pm$ and both crack-face integrals vanish. On $\Gamma$, $q = 1$ and $\boldsymbol{M} = -\bN$. Therefore,
\begin{equation}
\oint_{C} q\, H_{1J}\, M_J \, dC = -\int_{\Gamma} H_{1J}\, N_J \, d\Gamma = -J^\Gamma ,
\label{eq:jcontour6}
\end{equation}
i.e., introducing the weighting field closes the contour without changing the value of the integral. Applying the divergence theorem and the product rule gives
\begin{equation}
J^\Gamma = \int_{A_0} \left( P_{kJ} \frac{\partial u_k}{\partial X_1} - W \delta_{1J} \right) \frac{\partial q}{\partial X_J}\, dA_0
\;-\; \underbrace{\int_{A_0} q\, \frac{\partial H_{1J}}{\partial X_J}\, dA_0}_{R(q)} .
\label{eq:jcontour8}
\end{equation}
For a homogeneous hyperelastic body in equilibrium without body forces, i.e., with $W = W(\bF)$ having no explicit dependence on $\bX$ and $\bP = \partial W/\partial \bF$, a direct calculation shows that $\partial H_{1J}/\partial X_J = 0$ in $A_0$ \cite{eshelbyone, eshelbytensor}. The residual $R(q)$ therefore vanishes, and \eqn{eq:jcontour8} reduces to the domain form of the $J$-integral:
\begin{equation}
J^{A_0} = \int_{A_0} \left( P_{kJ} \frac{\partial u_k}{\partial X_1} - W \delta_{1J} \right) \frac{\partial q}{\partial X_J}\, dA_0 .
\label{eq:jcontour9}
\end{equation}
Although $q$ appears in the integrand, the value of the integral is the same for every continuous field $q$ that equals 1 on $\Gamma$ and 0 on $C_0$, since it always equals the contour integral $J^\Gamma$, as shown by \eqn{eq:jcontour8} with $R(q)=0$.

\subsection{The residual in numerical calculations}
\label{app:residual}

In FE or QC calculations with linear elements, $\bF^h$, $\bP^h$, $W^h$, and hence $\bH^h$ are constant within each element, so the constitutive conditions of \app{app:derivation} hold exactly within each element, and $\partial H^h_{1J}/\partial X_J = 0$ there. Here the subscript or superscript $h$ denotes quantities computed from the discrete fields, $h$ being the characteristic element size. Equilibrium, however, holds only in the weak (nodal-force) sense: the tractions, and hence $\bH^h$, jump across element faces. Consequently, $\partial H^h_{1J}/\partial X_J$ in \eqn{eq:jcontour8} does not vanish in distributional sense, and the residual is nonzero.
Writing \eqn{eq:jcontour8} for the discrete fields and rearranging gives
\begin{equation}
J_h^{A_0}(q) = J_h^{\Gamma} + R_h(q), \qquad R_h(q) = \int_{A_0} q\, \frac{\partial H^h_{1J}}{\partial X_J}\, dA_0 ,
\label{eq:jh_residual}
\end{equation}
where $J_h^{A_0}(q)$ is the domain integral of \eqn{eq:jcontour9} evaluated with the discrete fields, as computed in practice, and $J_h^{\Gamma}$ is the contour integral of \eqn{eq:jcontour2} along $\Gamma$ evaluated with the same fields; the latter does not depend on $q$. The residual $R_h(q)$ tends to zero as the mesh is refined and is the only term that depends on $q$. The uncertainty examined in \sect{sec:sampling} therefore originates from the discretization error of the fields in the integration domain.

\section{Detailed balance of $q$-field MCMC sampler}
\label{app:detailedbalance}

Following the Metropolis--Hastings framework, we now establish that the proposed MCMC sampling procedure over the space of admissible $q$-fields satisfies detailed balance. Let $\mathbf{q}=(q_1,\ldots,q_n)$ denote the interior nodal $q$-values, where $n = N_{\text{int}}$. The target distribution is the uniform distribution over the admissible set $[0,1]^n$,
\begin{equation}
p(\mathbf{q}) = 1 \quad \text{for } \mathbf{q}\in[0,1]^n,
\label{eq:target}
\end{equation}
corresponding to each interior $q$-value being independently and uniformly distributed on $[0,1]$. The proposal distribution, which selects one interior node uniformly at random and redraws its value from $\text{Uniform}(0,1)$ independently of the current state, is symmetric,
\begin{equation}
s(\mathbf{q}'\mid\mathbf{q}) = s(\mathbf{q}\mid\mathbf{q}').
\label{eq:symmetric}
\end{equation}
The Metropolis--Hastings ratio is therefore
\begin{equation}
r = \frac{p(\mathbf{q}')\,s(\mathbf{q}\mid\mathbf{q}')}{p(\mathbf{q})\,s(\mathbf{q}'\mid\mathbf{q})} = 1,
\label{eq:Qratio}
\end{equation}
since $p(\mathbf{q})=p(\mathbf{q}')=1$ from \eqn{eq:target} and the proposal terms cancel by \eqn{eq:symmetric}. The corresponding acceptance probability is
\begin{equation}
a(\mathbf{q}',\mathbf{q}) = \min(1,r) = 1,
\label{eq:acceptance}
\end{equation}
so every proposed update is always accepted, consistent with \alg{alg:mcmc_qfield}. The resulting transition kernel, $\kappa(\mathbf{q}'\mid\mathbf{q}) = s(\mathbf{q}'\mid\mathbf{q})\,a(\mathbf{q}',\mathbf{q})$, then satisfies detailed balance,
\begin{equation}
\kappa(\mathbf{q}'\mid\mathbf{q})\,p(\mathbf{q}) = \kappa(\mathbf{q}\mid\mathbf{q}')\,p(\mathbf{q}'),
\label{eq:detailedbalance}
\end{equation}
with respect to $p(\mathbf{q})$. Since the chain is additionally irreducible (any admissible $q$-field is reachable through successive single-node updates) and aperiodic (a continuous uniform draw returns to the same value with probability zero), it converges to $p(\mathbf{q})$, justifying the sampled $q$-fields as a valid statistical characterization of the admissible $q$-field space.

\section{Autocorrelation and Gelman--Rubin statistic of the sampled $J$-integral}
\label{app:acf}

\subsection{Autocorrelation}

Summing \eqn{eq:jele3} over the elements of the integration domain and collecting the terms of each node gives
\begin{equation}
J = \sum_e J^e = \sum_{\alpha} w^\alpha q^\alpha, \qquad
w^\alpha = -\frac{1}{B}\sum_{e \ni \alpha} H^e_{IK}\, N^\alpha_{,K}\, \hat n_I\, V^e ,
\label{eq:jlin_app}
\end{equation}
where the sum in $w^\alpha$ runs over the elements that contain node $\alpha$. The weights $w^\alpha$ depend only on the discrete fields and the mesh, and not on $q$. Since the contour nodes have the fixed values $q=1$ and $q=0$, their contribution is constant, and only the $N_{\text{int}}$ interior nodal values vary during sampling. In the stationary state, these values are independent, each with variance $v$, so that,
\begin{equation}
\operatorname{Var}(J) = \sum_{\alpha} (w^\alpha)^2 \operatorname{Var}(q^\alpha) = v \sum_{\alpha} (w^\alpha)^2 ,
\label{eq:varj_app}
\end{equation}
where the sums now run over the interior nodes.

Let $q^\alpha$ and ${q'}^\alpha$ be the values of node $\alpha$ at steps $m$ and $m+k$. Since different nodes are independent, $\operatorname{Cov}(q^\alpha, {q'}^\beta) = 0$ for $\alpha \neq \beta$, and
\begin{equation}
\operatorname{Cov}(J_m, J_{m+k}) = \sum_\alpha (w^\alpha)^2 \operatorname{Cov}(q^\alpha, {q'}^\alpha) .
\label{eq:covj_app}
\end{equation}
If node $\alpha$ is not selected during the $k$ steps, which occurs with probability $(1-1/N_{\text{int}})^k$, then ${q'}^\alpha = q^\alpha$ and $\operatorname{Cov}(q^\alpha, {q'}^\alpha) = v$. Otherwise, ${q'}^\alpha$ is a new independent draw and $\operatorname{Cov}(q^\alpha, {q'}^\alpha) = 0$. Hence
\begin{equation}
\operatorname{Cov}(q^\alpha, {q'}^\alpha) = v \left(1 - \frac{1}{N_{\text{int}}}\right)^{k},
\end{equation}
and substituting into \eqn{eq:covj_app} and using \eqn{eq:varj_app} gives,
\begin{equation}
\operatorname{Cov}(J_m, J_{m+k}) = \left(1 - \frac{1}{N_{\text{int}}}\right)^{k} v \sum_\alpha (w^\alpha)^2 = \left(1 - \frac{1}{N_{\text{int}}}\right)^{k} \operatorname{Var}(J).
\end{equation}
Dividing by $\operatorname{Var}(J)$ gives the autocorrelation
\begin{equation}
\rho(k) = \frac{\operatorname{Cov}(J_m, J_{m+k})}{\operatorname{Var}(J)} = \left(1 - \frac{1}{N_{\text{int}}}\right)^{k},
\end{equation}
which is independent of the weights $w^\alpha$, and hence of the fields and the load. With $\lambda = 1 - 1/N_{\text{int}}$, the integrated autocorrelation time follows from the geometric series,
\begin{equation}
\tau = 1 + 2\sum_{k=1}^{\infty} \lambda^k = 1 + \frac{2\lambda}{1-\lambda} = 2N_{\text{int}} - 1 .
\end{equation}

\subsection{Gelman--Rubin statistic}

Let $J_{ij}$ be the $i$-th post-burn-in sample of chain $j$, with $N_{\text{sample}}$ samples in each of $N_c$ chains. With the chain means $\bar J_j$, the overall mean $\bar J$, and the chain variances $s_j^2$,
\begin{equation}
\bar J_j = \frac{1}{N_{\text{sample}}}\sum_{i} J_{ij}, \qquad
\bar J = \frac{1}{N_c}\sum_{j} \bar J_j, \qquad
s_j^2 = \frac{1}{N_{\text{sample}}-1}\sum_{i} \big(J_{ij} - \bar J_j\big)^2 ,
\end{equation}
the within-chain variance $V_{within}$ and the between-chain variance $V_{between}$ are
\begin{equation}
V_{within} = \frac{1}{N_c}\sum_{j} s_j^2, \qquad
V_{between} = \frac{N_{\text{sample}}}{N_c - 1}\sum_{j} \big(\bar J_j - \bar J\big)^2 ,
\end{equation}
The pooled variance $\hat V$ and the Gelman--Rubin statistic $\hat R$ are then ~\cite{gelman:rubin:1992}
\begin{equation}
\hat V = \frac{N_{\text{sample}}-1}{N_{\text{sample}}}\, V_{within} + \frac{V_{between}}{N_{\text{sample}}}, \qquad
\hat R = \sqrt{\frac{\hat V}{V_{within}}} .
\end{equation}

\newpage
\bibliographystyle{unsrt}
\bibliography{references}

\end{document}